# A Complex Systems Approach to Exoplanet Atmospheric Chemistry: New Prospects for Ruling Out the Possibility of Alien Life-As-We-Know-It

Theresa Fisher[1], Estelle Janin[1], Sara Imari Walker[1,2,3*]
[1]School of Earth and Space Exploration, Arizona State University, Tempe AZ USA
[2]Beyond Center for Fundamental Concepts in Science, Arizona State University, Tempe AZ USA
[3]Santa Fe Institute, Santa Fe, NM USA
*author for correspondence: sara.i.walker@asu.edu

## Abstract

The near-term capability to characterize terrestrial exoplanet atmospheres may bring us closer to discovering alien life through atmospheric data. However, remotely detectable candidate biosignature gases are subject to false positive signals because they can also be produced abiotically, raising a critical need to develop methods to determine whether a gas is produced abiotically or not. To distinguish biological, abiotic and anomalous sources (unidentified as abiotic or biotic) of biosignature gases, we take a complex systems approach implementing chemical reaction network analysis of planetary atmospheres. We simulated 30,000 terrestrial atmospheres, organized in two datasets: Archean Earth-like worlds and modern Earth-like worlds. For Archean Earth-like worlds we study cases where $CH_4$ is produced abiotically via serpentinization, biologically via methanogenesis, or from anomalous sources. We also simulate modern Earth-like atmospheres with and without industrial CFC-12. We show how network properties can effectively distinguish scenarios where $CH_4$ is produced from methanogenesis and serpentinization. Network analysis also distinguishes modern Earth-like atmospheres with CFC-12 from those without it. Using Bayesian analysis, we demonstrate how atmospheric network statistics can provide stronger confidence for ruling out biological explanations compared to gas abundance statistics alone. Our results confirm how a network theoretic approach allows distinguishing hypotheses about biological, abiotic and anomalous atmospheric drivers, and importantly, allows ruling out life-as-we-know-it as a possible explanation. We conclude with a discussion of how further developing statistical inference methods for spectral data that incorporate network properties could significantly strengthen future biosignature detection efforts.

## Main

The James Webb Space Telescope (JWST) has already revolutionized our understanding of terrestrial exoplanet atmospheres, with important recent observations of TRAPPIST-1 planets suggesting potential atmospheric loss on some worlds (Greene et al. 2023) while leaving open questions about others (Zieba et al. 2023). These early results, combined with detailed atmospheric modeling efforts (Lincowski et al. 2023; Moran et al. 2023; Meadows et al. 2023), demonstrate both the power and limitations of current telescope capabilities. The imminent operation of ground-based observatories capable of high-resolution spectroscopy, such as the Extremely Large Telescope (ELT), along with direct-imaging telescopes like the future Habitable Worlds Observatory (HWO), will further expand our ability to characterize terrestrial exoplanet atmospheres. As we enter this new era of exoplanet characterization, developing robust methods to analyze atmospheric data and test hypotheses about abiotic or biological explanations, becomes increasingly crucial. For robust life detection to become possible we must be able to identify the most effective method(s) for leveraging data from these observatories.

Early approaches to exoplanet life detection relied on the identification of 'smoking guns', once viewed as indisputable evidence for life (Meadows 2017). The idea of smoking guns was typically associated to observationally accessible gaseous molecules that are produced in abundance through the evolution of life on Earth. An example is $O_2$ (Tremblay et al. 2019), particularly in concert with $CH_4$ (Tremblay et al. 2019; Kaltenegger, Lin, and Rugheimer 2020; Des Marais et al. 2002; Sagan et al. 1993; Hitchcock & Lovelock 1967), which became abundant in Earth's atmosphere through the evolution of oxygenic photosynthesis and methanogenesis, respectively. However, as knowledge of exoplanets increased and the immense diversity of worlds became apparent, atmospheric molecular biosignatures – most prevalent among them $O_2$ – have come under more scrutiny due to the risks for false positives (Wordsworth and Pierrehumbert 2014; Domagal-Goldman et al. 2014; Meadows et al. 2018; Harman and Domagal-Goldman 2018) and false negatives (Reinhard et al. 2017). Methane can also be ambiguous (Guzmán-Marmolejo, Segura, and Escobar-Briones 2013). Challenging cases include the well-studied $O_2$-$CH_4$ chemical disequilibrium, which can, for example, be mimicked if the atmosphere of an exoplanet's unseen moon contaminates the spectra of its host exoplanet, leading to the misleading appearance of a single body with a strong atmospheric disequilibrium (Rein, Fujii, and Spiegel 2014). In short, there are many hypotheses, abiotic and biotic, that could explain a given gas's atmospheric abundance and we lack tools to distinguish abiotic from biotic explanations (Smith & Mathis 2023; Foote et al. 2023).

The atmospheric composition of exoplanets is also a target in the search for extraterrestrial technology (Wright 2018; Haqq-Misra et al. 2022). Several industrial gases have been posited as technosignatures, including $NO_2$ (Kopparapu, Arney, and Haqq-Misra 2021), as well as overall changes in the mixing of nitrogenous gases (Haqq-Misra and Fauchez 2022). Chlorofluorocarbons (CFCs) are also of interest, due to their distinctive spectral signatures (Kopparapu, Arney, and Haqq-Misra 2021; Haqq-Misra, Kopparapu, et al. 2022) and expected longevity once introduced into an atmosphere (Balbi and Ćirković 2021). CFCs have been posited as robust technosignatures because no other known processes, whether biological or geological, have been identified that can produce these compounds, with all known formation mechanisms requiring specific industrial syntheses of complex carbon-chlorine-fluorine bonds (Seinfeld & Pandis, 2006). However, other species of halocarbons can be produced by volcanic activity (Visscher et al. 2004; Broadley et al. 2018; Klobas and Wilmouth 2019). Note that CFC spectral detection is only barely within near-term capabilities; it is estimated that CFC emissions would have to be an order of magnitude, or more, greater than the peak of human CFC emissions to be readily detectable by JWST (Lin, Abad, and Loeb 2014; Haqq-Misra et al. 2022). Currently most atmospheric technosignature research focuses on gases like CFCs that are industrial byproducts; however, as the community moves towards more general methods for technosignature detection, a future goal might be to detect technosignatures produced by unknown technological processes, which may or may not include atmospheric gases that we recognize from Earth's technology.

Another challenge is that alien worlds might harbor non-Earth-like atmospheres, leading to different or unexpected atmospheric biosignatures (Schwieterman et al 2018). For example, $CH_3Cl$ has been proposed as a biosignature for otherwise Earth-like and $O_2$-rich atmospheres found on planets orbiting M dwarf stars, whose UV spectral energy distributions may allow for greater accumulation of this gas (Segura et al. 2005; Gebauer et al. 2021). Further alternative proposal for atmospheric biosignatures include $PH_3$ for anoxic worlds (Sousa-Silva et al. 2019), $NH_3$ and $CH_3Cl$ for hydrogen-dominated atmospheres (Seager, Bains, and Hu 2013a; Bains, Seager, and Zsom 2014; Wunderlich et al. 2020), and halocarbons in biospheres using chlorinic photosynthesis (Haas 2010). Each assumes a different hypothesis for the dominant metabolism(s) that could persist on an alien world, and thereby carries different possible false positive scenarios. Even Earth's own biosignature gases were considerably different from the modern atmosphere during the bulk of Earth's history. In particular, Earth's $O_2$ biosignature has only been detectable for the past 1/8th of its inhabited history due to negligible atmospheric $O_2$ levels during the Archean and still very low levels – arguably too low to be detectable – until 0.75 Gy ago (Wogan & Catling, 2019; Arney et al. 2016; Reinhard et al. 2017). This also holds for atmospheric technosignatures, which are even less constrained and must be understood in the context of their surrounding biosphere (Haqq-Misra et al. 2022). Technosignatures are especially challenging given that these have only arisen within the last 50-100 years, a tiny sliver of Earth's history.

The observational challenges and confounding abiotic factors confronting exoplanet life detection efforts suggest a need for new approaches (Walker et al. 2018; Sandora & Silk 2023) with some researchers advocating searching for anomalies (Cleland 2019; Kinney & Kempes 2022; Sarkar et al. 2022) in the form of signals that *cannot* be explained by a known abiotic or biological mechanism. Anomalies are signals without explanation, which might be sources by unknown abiotic or biological processes. In our prior work, we showed how, over a wide range of temperatures, network metrics could be reliable indicators of the distance from thermochemical equilibria in hot Jupiter atmospheres (Fisher et al. 2022). Building on this, in the current work we now evaluate the effectiveness of network-based approaches in identifying features of atmospheres driven by known biological or technological processes, with the aim to establish methods to identify what is distinct about these atmospheres at a systems-level from those driven by known abiotic processes, and those that are anomalous (not confirmed as a known abiotic, biological or technological process). Given recent emphasis on identifying anomalies as useful in the search for alien life, it is important to determine whether we can distinguish genuine anomalies from known biological sources (Cleland 2019; Kinney and Kempes 2022). The goal is not necessarily to determine whether we can confirm that life is present, but also to potentially rule out concrete hypotheses that are not consistent with the data (Foote et al. 2023) about *what kind of life could be present*. In some cases, this may include ruling out the hypothesis that life, if it exists on a particular exoplanet, uses the same metabolisms or technological processes as we know on Earth.

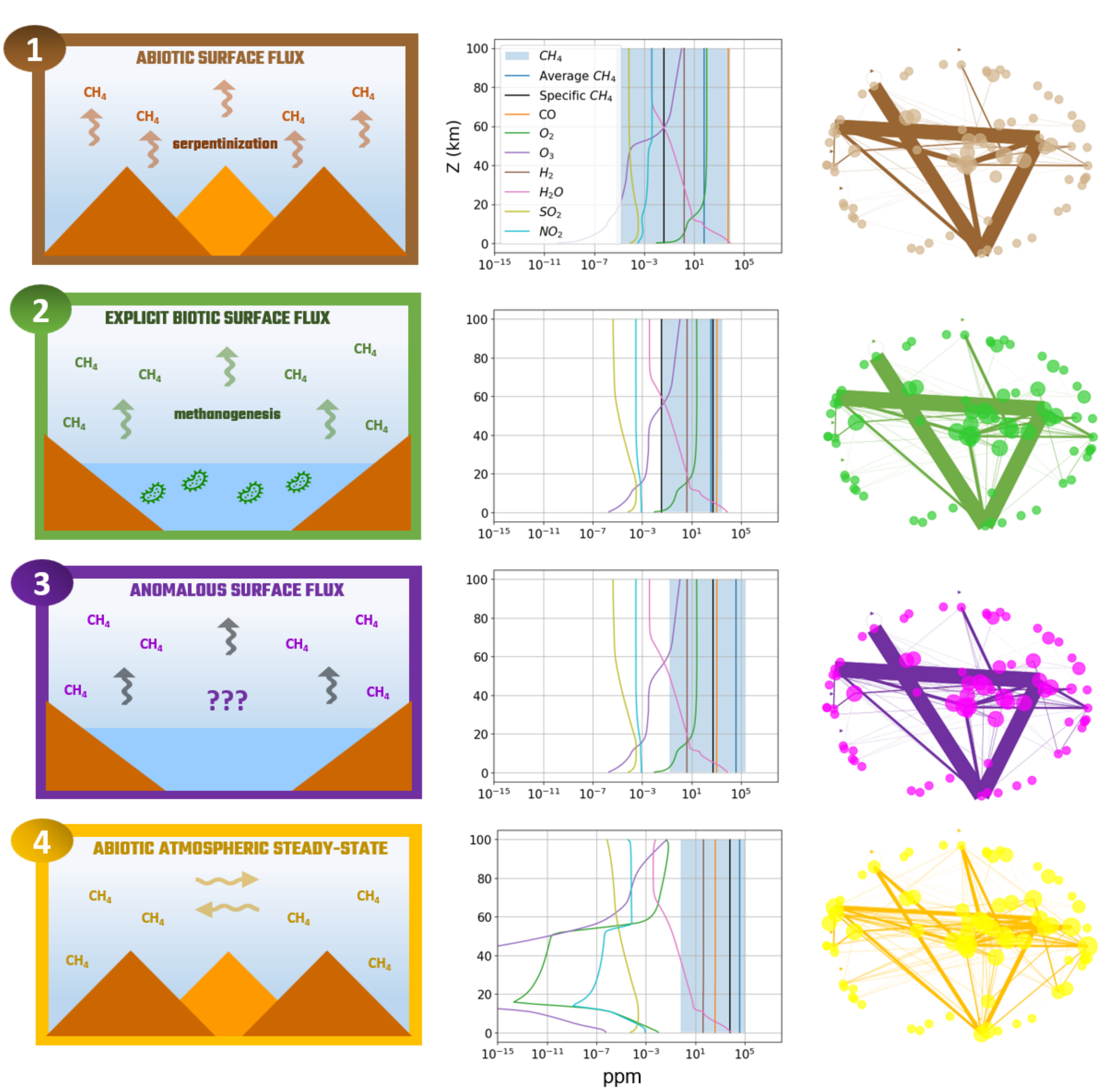

***Figure 1:*** *Conceptual representations, abundance profiles, and network visualizations for each category of Archean Earth atmospheric model used in this study. The first column shows schematics of the scenarios under study. The second column shows the vertical profiles of the atmospheric species abundances corresponding to the initial conditions for each scenario. All initial abundances are fixed, except for methane. The blue shaded window corresponds to the range of methane abundances of simulations of this scenario, the solid blue line corresponds to the average methane concentration for all the simulations and the solid black line corresponds to the methane vertical profile for one example instance, randomly extracted from the simulation batch. For the network visualizations, the nodes are not labeled for clarity but in the case of the first three scenarios, the three dominant edges connect the following nodes: NO, $NO_2$, $HO_2$ and SO2ii (in an excited state) – i.e., these are very reactive species in these atmospheres. From top to bottom we show the following scenarios studied herein: 1)* ***Abiotic flux from serpentinization.*** *Methane is modeled with a known abiotic flux mechanism, sourced from the planetary surface as the product of serpentinization. 2)* ***Biotic surface flux.*** *Methane is modeled with a known biological surface flux mechanism like that expected to have been present on the Archean Earth, with methanogenesis reactions from Earth's metabolisms explicitly included in the model. 3)* ***Anomalous surface flux.*** *Methane is modeled as the result of a surface flux of undetermined origin. Flux rates are set to be comparable to that of biotic production rates on Earth (biotic surface flux model). 4)* ***Anomalous atmospheric steady state.*** *Methane is modeled as a component of a steady state atmosphere, with abundance levels comparable to that produced by biology on Earth (biotic surface flux model).*

# Modeling Atmospheres as Complex Systems

Early research applying complex systems approaches to planetary atmospheres by Solé and Munteanu (2004) constructed networks from the characterized atmospheric chemistry of all major atmospheres in the Solar System. They found Earth's atmosphere to display a noticeably higher degree of organization and exhibit hierarchical and modular properties not seen in the networks of other Solar System atmospheres; these findings were also confirmed in a more recent, and quantitatively similar analysis by Wong et al (Wong et al., 2023). However, it should be noted that most of the other atmospheres used in these studies, beyond that of the Earth, lack sufficient knowledge of the diversity of chemical species to construct networks with enough species and reactions to yield statistically meaningful results for many of the network properties used, like the degree distribution (see e.g., Broido and Clauset 2019 for the challenges of fitting distributions to small datasets). Focusing on Earth's atmospheric reaction network, which has much more data and is already known to be driven by life, has also demonstrated how network representations of the chemistry can exhibit emergent features, such as algebraic closure (Centler and Dittrich 2007) and high returnability (Estrada 2012) that are not characteristic of random networks and therefore could be specific to Earth as a living world. Similar network topological features are also seen at all scales of the Earth's biochemistry (Wuchty, Ravasz, and Barabási 2006; Kim et al. 2019), from cellular biochemical networks (Barabási and Oltvai 2004; Daniels et al. 2018) to individual organismal metabolic networks (Jeong et al. 2000), to whole ecosystems (Solé et al. 2002; Marquet et al. 2005) and the entirety of cataloged reactions catalyzed by Earth's modern biosphere (Kim et al., 2019). These studies show how biochemical networks can be distinguished from non-biological networks through specific motifs (subgraphs within the network) and topological scaling laws, which taken together lend support to the idea that network-based approaches might provide new tools for agnostic life detection. However, direct comparison of network representations of data collected across different systems remains a challenge. This is because the systems of study – e.g., the biochemistry of *E. coli* versus the chemistry of the interstellar medium (Jolley & Douglas 2011) or even Venus and Earth in Wong et al. (Wong et al., 2023) – involve data collected from very different sources, and under different physical and chemical constraints, leading in some cases to rather substantial differences in the physical interpretation of the network representations across examples under study. Furthermore, many kinds of projections of data into a mathematical representation as a network graph are possible; each of them has subtly different interpretations with respect to their physical meaning and predictive capacity, and these different representations of the same data can lead to very different network statistics (Montañez et al 2010, Smith et al 2019) and therefore different results when contrasting network properties across systems.

These challenges indicate that it is highly non-trivial to (1) construct a physically and/or chemically meaningful graphical representation of a given complex system such as a planetary atmosphere. And (2) a standardized approach is necessary for both generating network representations of planetary atmospheres and cross-comparing features consistent with living versus non-living worlds to confirm or refute specific hypotheses for explaining atmospheric data. These challenges need to be overcome if the potential promise of network-theoretic approaches to life detection in exoplanet atmospheres is to be realized. Herein, our models are therefore constructed to be directly comparable because they are built under the same sets of constraints and assumptions and from data that is prepared, treated and analyzed the same way. By doing so, we aim to provide a rigorous first test of how network theory might provide a useful quantitative approach to discriminating atmospheres driven by known abiotic or biological mechanisms, from any of the variety of unknown sources that may be at play in exoplanet data.

In this work, we primarily focus on analyzing Archean Earth-like atmospheres to evaluate the effectiveness of network-based approaches in identifying features driven by known biological processes (methanogenesis) as distinct from those driven by known abiotic processes (serpentinization) and those that are anomalous. We then extend this analysis to a case study of Modern Earth-like atmospheres and the production of CFCs by technology to demonstrate the broader applicability of our approach. Both case studies were chosen to reflect standard scenarios for alien life used on exoplanet biosignature literature (Arney et al 2016, Haqq-Misra et al 2022).

# Archean Earth Models and Methods

We aimed to compare case studies based on Earth and its evolved biology, to case studies that are intentionally set up as anomalous signals producing similar abundance profiles with the goal to determine if network metrics can rule out hypotheses to explain signals based on known mechanisms. Methane was chosen as our target biosignature gas for this study, both due to its important role in defining the atmospheric composition of Archean Earth (Kasting and Siefert 2002) and its potential spectral detectability (Arney et al 2016). We built statistics based on ensembles of Archean Earth-like atmospheres classified in four main scenarios, with one known abiotic mechanism (serpentinization), one known biotic mechanism (methanogenesis), and two cases where methane is of unknown origin (anomalous) designed to match signals produced by the known biotic mechanism. The models are shown in **Figure 1**, and described in detail below:

1. **Abiotic surface flux from serpentinization (Figure 1, top row).** We simulate methane production as an abiotic surface flux driven by serpentinization (Thompson et al. 2022), thereby modeling a known abiotic production of methane on Earth. We include this model set as a potential abiotic false-positive case for comparison to the model with a biological surface flux. Methane flux values were randomly sampled from a Gaussian distribution centered around $10^6$ molecules/cm$^2$, with a standard deviation of $10^2$, and a range between $10^{-2}$ and $10^8$ molecules/cm$^2$, as constructed from the values given in Thompson et al (2022). Note that the high flux limit is a cutoff value not affecting the symmetry of the Gaussian distribution.
2. **Explicit biotic surface flux (Figure 1, second row).** Methane production is simulated as a biological surface flux, and a methanogenesis reaction derived from Earth's metabolisms was explicitly included in the reaction network (Catling and Zahnle 2020). This scenario models a true positive case for life detection. Methane flux values are drawn from a Gaussian distribution centered around $10^{11}$ molecules/cm$^2$, with a standard deviation of $10^2$, and a range between $10^3$ molecules/cm$^2$ and $10^{13}$ molecules/cm$^2$ (cutoff value), as constructed from values given in Arney et al. (2016) consistent with methanogenesis on Earth.
3. **Anomalous surface flux (high flux case, Figure 1, third row).** Methane is modeled as a surface flux with the same flux levels as the biotic surface flux scenario, but the reaction network is not explicitly coupled to a biologically derived methanogenesis reaction. This model set is referred to as an *anomalous surface flux* since the source of the flux is not specified. This set of simulations was included to present a case study in distinguishing between an anomalous source and methanogenesis akin to Earth biology (described by the biotic surface flux model scenario above). Methane flux values are drawn from a Gaussian distribution centered around $10^{11}$ molecules/cm$^2$, with a range between $10^3$ molecules/cm$^2$ and $10^{13}$ molecules/cm$^2$, consistent with sampling from the same distribution of flux values as implemented in the biotic surface flux model. Note that the anomalous surface flux scenario could be either associated with life or an abiotic process, and while the mechanism is not specified the flux rate is the same as what is observed in methanogenesis on Earth.

4. **Anomalous abiotic atmospheric steady state (low flux case, Figure 1, bottom row).** Methane abundance was modeled as part of an atmospheric steady state, with the same abundances as the biotic model scenario, i.e., with identical $CH_4$ mixing ratios, but not derived from a surface flux. This represents another possible anomalous scenario for false positives, where the methane concentration ranges from $10^{-4}$ to $10^4$ ppm with the value being held constant through an unspecified mechanism in the model. This scenario is probably the least realistic based on current knowledge because atmospheres with a fixed mixing ratio of $CH_4$ rarely exist in nature, and atmospheres with $CH_4$-rich secondary atmospheres are difficult to produce outside of a biological context (Tian and Heng 2023). We include it because of its utility as an additional control, allowing us to test how well network statistics can distinguish this case from a known biological mechanism.

The model sets implemented in this study are designed as well-defined, simplified case studies to allow assessing how distinguishable different known abiotic and biotic scenarios could be as compared to anomalous explanations. We include anomalous signals in our analyses to open discussion on scenarios for the possibility of life-as-we-don't-know-it or even geochemistry-as-we-don't-know-it, which would have unknown metabolism, as a possible explanation for a given observation. Even though they may not all represent realistic exoplanet scenarios, they allow us to isolate how atmospheric network properties are influenced by different potential abiotic and biotic surface fluxes and how we might distinguish these cases from anomalous cases where a potential detected gas is not explained in terms of known biological or abiotic processes.

For each of the above modeled scenarios, we generated an ensemble of 5,000 atmospheric simulations. Each simulation was run on 200 vertical layers with the atmospheric code `atmos`, using only the photochemical component and a fixed temperature-pressure structure (see **Supplement Figure S1**) to converge on consistent solutions for the chemical profiles (including aerosols). The second column of **Figure 1** shows the vertical profiles of the atmospheric species abundances corresponding to the initial conditions for each scenario. Note we do not self-consistently vary surface boundary conditions between each scenario; for example, in a truly abiotic case, the CO deposition velocity should be about four orders of magnitude smaller than in the biotic case, see Kharecha et al 2005. Instead, all initial abundances are fixed, except for that of methane, allowing us to better isolate the impact of specific changes and hypotheses on network metrics, without accounting for the variety of other factors in more realistic scenarios that would also impact network analyses, which can be a subject of future studies aiming to better refine analyses to specific planetary scenarios once the foundations of the methods of network analyses and its sensitivities are fledged out.

To simulate such many atmospheres, we used the `PyAtmos` Python wrapper to run batches of `atmos` models in parallel (Chopra et al 2023), see Supplement Section "Generating Ensembles of Simulated Exoplanet Atmospheres with `atmos`" for details. Finally, to make this study more relevant to the large population of small terrestrial planets detected around red dwarfs, we simulated these atmospheres as if the planet were orbiting the star GJ 581 (M5 V) and used the corresponding stellar spectrum from the MUSCLES Survey (France et al. 2016). We then scaled the semi-major axis of the planet such that it receives the same amount of stellar flux as the Earth around the Sun.

Each modeled atmosphere in each dataset was used to generate a chemical reaction network (CRN) of the chemistry in the atmosphere, where species are connected if they participate in a common reaction, and edges are weighted according to the modeled reaction rates (see Section "Constructing and Measuring Atmospheric Reaction Networks" and **Supplement Figure S4** for details). The third column of **Figure 1** shows graphical representations of the CRN for each scenario, built using the median value of the $CH_4$ abundance window from initial conditions (thus close to the average network for each case). The resulting network for the Archean Earth group has 74 nodes and 704 edges for the abiotic and anomalous cases, and 77 nodes and 711 edges for the biotic case. The additional edges in the biotic case arise due to the incorporation of biologically driven reactions, where the additional nodes are fictitious species added to `atmos` to model methanogenesis (see Supplement Section "Archean Earth-like Worlds with and without Biologically Produced $CH_4$"). These species are short-lived and do not yield noticeable impact on network statistics (see **Supplement Figures S2** and **S3**). Species abundances used to calculate reaction fluxes and edge weights were taken at a geopotential height of z = 63.8 km, which corresponds to a pressure of about 0.2 mbar, as this represents a slice of the atmosphere that is likely to be accessible to remote detection while having sufficient molecular abundances for chemical network analysis. We also ran the same analysis at z = 47 km, corresponding to a pressure of 1 mbar, and found no significant difference in the results of our analysis (see Supplement Section "Comparison between 0.2 mbar and 1 mbar Probed Altitudes", **Figures S13-15**).

From these reaction rate-weighted networks, we calculated several different network measures for each atmosphere, including mean degree, clustering coefficient, node betweenness centrality, and average nearest neighbor degree. See **Table 1** for a description of these quantities and their physical interpretations. We used these data to generate distributions of network measures for each model scenario built over all 5,000 atmospheres in each simulated ensemble. We also collected data from the simulations to generate distributions of the abundances of key atmospheric gases, here chosen to be our candidate biosignature gas, i.e., $CH_4$. The resulting distributions were then compared across model sets, using the Anderson-Darling (tail-weighted Cramer-von-Mises) k-sample test, which allows to determine whether two samples of data are drawn from the same population with a wide variety of differences in samples (spread, tails, etc.). The higher the Anderson-Darling (AD) value, the better the network measure (or gas abundance) is at distinguishing between different model scenarios, with a threshold corresponding to a p-value < 0.01.

| **Symbol** | **Network measure** | **Description in the context of CRNs** | **Physical interpretation** |
| --- | --- | --- | --- |
| $<k>$ | Mean degree | Average of the total flux driven by each chemical compound across the network | Higher weighted mean degree indicates networks where species participate in more reactions |
| $<k_{nn}>$ | Average of average neighbor degree | Average of the total flux driven by the neighbors of each chemical compound across the network | High average neighbor degree indicates that reactive species co-participate in many reactions |
| $<l>$ | Average shortest path length | Average of the minimum amounts of flux along dependency pathway between every pair of compounds | Fewest reactions required to convert one species to another in a given direction, as weighted by the flux |
| $<C>$ | Average clustering coefficient | Average tendency across the network of a chemical compound to have neighbors that participate in common chemical reactions | How often species co-participate in reactions with the shared sets of reactants |
| $<g(v)>$ | Average node betweenness centrality | Average tendency of a compound to be part of the shortest (flux-weighted) reaction path between every pair of compounds | Whether some species play a more central role in multiple reaction pathways instead of a uniform influence of all species |
| $<g(u)>$ | Average edge betweenness centrality | Average tendency of a reaction to be part of the shortest (flux-weighted) reaction path between every pair of compounds | Whether some reactions play a more central role in multiple reaction pathways instead of a uniform influence of all reactions |

***Table 1.*** *Network measures used in this study with their definitions, notation and physical interpretation as a statistical measure of planetary atmospheric chemistry.*

# Results

## Comparing Known Abiotic and Biotic Sources of Methane to its Production by Anomalous Sources in Archean Earth-like Atmospheres

We apply the Anderson-Darling test to compare pairs of distributions generated from the different scenarios for Archean Earth-like atmospheres, using the case of a known biological production mechanism (methanogenesis) as the reference distribution for comparison with the other cases for both the network metrics and species abundances comparisons. We use the non-parametric k-sample version of the test, which assesses if two or more sets of data were drawn from the same distribution (Schloz & Stephens, 1978). While other conventional statistical tests (e.g., the Kolmogorov-Smirnov 2-sample test) are only sensitive to differences in medians, the Anderson-Darling test is sensitive to a wider variety of differences between distributions (spread, tails, etc.). We chose a critical value equivalent to a 0.1% significance level, i.e., 6.546 (corresponding to the horizontal dotted line in the **Figure 2** and **Figure 5**) above which a pair of given scenarios can be confidently distinguished from one another. In addition, to compare the network representations to another system-level but non-network-based metric, we also computed the variance of the distributions of species abundances for each modeled scenario. That is, we rank ordered all the species by abundance in each simulation output and distilled that curve down to its variance. This allows controlling for not having a strong underlying model of where the observed species came from, while probing features of the distribution of species in the atmosphere but also not explicitly using a network representation of the species present.

When comparing data across the four Archean Earth-like scenarios, we find that each network metric is independently able to distinguish data sampled from abiotic and anomalous cases as distinct from data sampled from the ensemble of atmospheres with known biology, see **Figure 2**. A higher Anderson value indicates a greater dissimilarity between the compared distributions, giving more confidence that we can reject the null hypothesis that the two distributions were sampled from the same source. Methane abundance and the variance in the species abundances are good discriminators of known abiotic and biotic sources for its production: the biotic scenario yields larger methane concentrations, on the order of a mixing ratio of $10^4$ ppm, whereas abiotic concentrations did not exceed $10^3$ ppm (top row **Figure 2**). This is consistent with the use of $CH_4$ abundance as a biosignature gas for known methane-producing metabolisms on Archean Earth. However, when comparing the biological case of methane production to either of the anomalous cases, we find the $CH_4$ abundance and species distribution variances are still distinguishable but the lower Anderson values suggest more similarity between these cases.

When the known abiotic distribution was compared to the known biological distribution, neither $CH_4$ gas abundance nor the variance in the species abundance distributions was the most dissimilar feature. Mean degree $<k>$ and the average shortest path length $<l>$ were more dissimilar (see **Figure 2**, rows 3 and 4). This indicates that network measures may pick up more significantly on the differences between the different scenarios than $CH_4$ gas alone does. Furthermore, in cases comparing known biological to anomalous cases, we find that network measures provide greater dissimilarity to distinguish these cases than $CH_4$ gas abundances do, because the biological and anomalous $CH_4$ abundance distributions are not so dissimilar (lower Anderson values in bottom panel of **Figure 2**). Mean degree $<k>$, average node betweenness centrality $<g(v)>$, and average shortest path length $<l>$, exhibit more dissimilarity when comparing biological to anomalous cases than what we see for $CH_4$ abundance. Each of these network measures captures facets of the behavior of highly reactive species, like $CH_4$, that are important participants in multiple chemical reactions. Thus, it appears that the network measures are more efficient at picking up the system-level differences in the atmosphere related to the different mechanisms for introducing $CH_4$. In particular, the much higher Anderson values for distinguishing between the biological and the anomalous steady state scenarios using network measures suggest a high utility for situations where there may be anomalous sources of methane on an exoplanet that cannot be built into known flux sources in an atmospheric model.

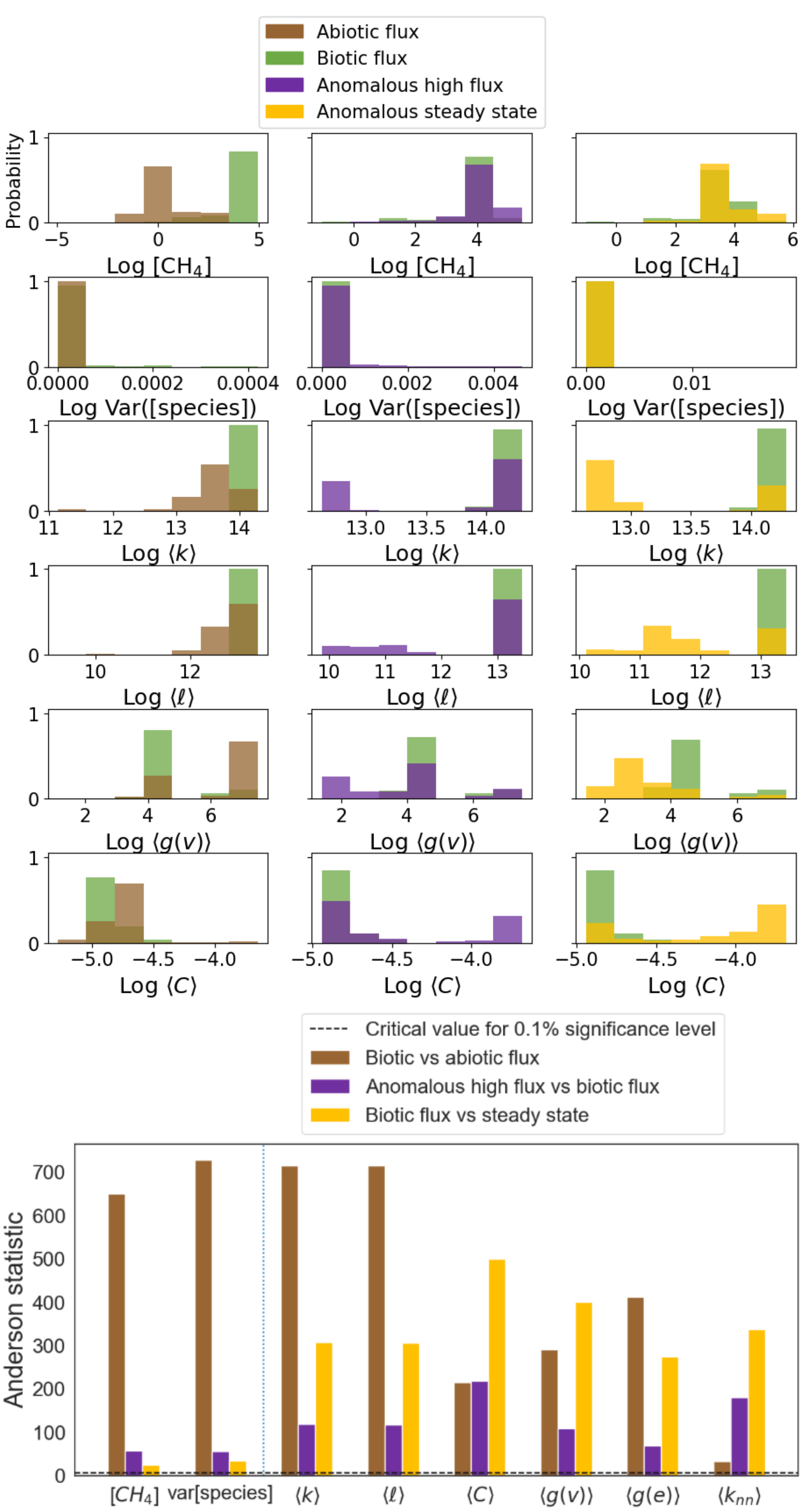

***Figure 2.*** *Top: Distributions of the probability (y-axis) of a simulated atmosphere yielding the given network metric value, variance in the distribution of species abundances or $CH_4$ abundance in log ppm (x-axis) over each of the sampled Archean Earth datasets, probed at 0.2 mbar. Scenarios included are: models where methane is modeled as an abiotic surface flux (brown); models where methane is modeled as a biotic surface flux with the methanogenesis pathway explicitly included in the chemical reaction network (green); models where methane was modeled as originating from a surface flux of undetermined origin, e.g., anomalous surface flux (purple); models where methane is modeled as an anomalous steady-state component of the planet's atmosphere with unknown source (yellow). Bottom: Anderson scores of the biotic flux scenario as compared to the three control models (abiotic flux, anomalous surface flux, and anomalous steady state). Each bar indicates the performance of the corresponding metric in distinguishing between different modeled scenarios. The horizontal black dotted line indicates the critical value (6.546) beyond which scenarios can be distinguished at 0.1% confidence level. The vertical blue dotted line separates $CH_4$ concentration and species abundance variance from the network metrics.*

In many cases we should expect that we will not know all atmospheric gases and reactions present in each exoplanetary atmosphere. We therefore performed perturbative tests to simulate cases where we might not have full knowledge of atmospheric composition. We re-ran our analyses with $C_2H_2$ removed from the simulations of known abiotic and biological cases, see **Supplement Figure S5**. The species $C_2H_2$ is known to play a key role in the formation of hazes: eliminating it removes a (poorly constrained) hydrocarbon sink. This significantly changes the behavior of the chemistry of the atmosphere by producing much higher levels of long-chain hydrocarbons. In these experiments, we found that network measures performed better than methane abundance distributions at distinguishing between the perturbed simulations and the original abiotic and biological datasets (see **Supplement Figure S5**). This provides further confirmation that network measures are more sensitive to system-level differences than gas abundances alone, indicating that if network properties could be directly inferred from exoplanet atmospheric data, they could provide a more sensitive window into differences in large scale features, potentially even those associated with life. In addition, we tested the robustness of network properties after the removal of a node with similar chemical properties and the same degree as methane for the Archean Earth cases: $CH_3CO$. We found no significant impact on global network statistics (not shown), indicating we are indeed capturing network-wide effects of different methane-associated processes involved in this study (abiotic and biotic), but further analyses of a variety of species in the network and studies of the impact of their inclusion or removal would be necessary to fully determine how different chemical processes impact network structure.

## Inferring the Presence of Life from Network Data

The preceding examples demonstrate how network structure can distinguish atmospheres driven by biological surface fluxes from both abiotic and anomalous examples, even in cases where the biosignature gas driving the differences is itself not as good a discriminator. A next question is whether network theory could be used to infer the presence of life, and how robust this inference will be depending on how common or rare life is.

Bayesian approaches have been adopted in several recent life detection studies due to their utility in quantitatively assessing sensitivity of candidate biosignature signals to the prior probability for life (Walker 2017, Walker et al 2018, Catling et al 2018, Affholder et al. 2021, Foote et al 2023, Smith & Mathis 2023). The prior probability for life, *P(life)*, i.e., the probability that life might exist on other planetary bodies, is currently unconstrained on the interval (0,1]: we know it is not exactly zero because life emerged on at least one planet (Earth), but we do not know how many others harbor life. This probability is therefore currently unconstrained and will continue to be unconstrained until we make a detection of alien life, rule out the possibility of life on some habitable worlds, or the mechanism for the origin of life on a planetary body is solved (Walker 2017). In fact, a primary, but often unstated goal, of exoplanet science is to constrain this probability by making a positive detection of alien life or at least ruling out some of the parameter space. Many biosignature candidates are sensitive to the value of *P(life),* including atmospheric gases that have abiotic mechanisms for their production: if there is a false positive scenario for a given biosignature, tighter constraints on *P(life)* are required *a priori* to confirm biogenicity (Walker et al 2018). This suggests that the exoplanet community could adopt a strategy to target biosignature candidates that are robust to large variation in the value *P(life)*, and also are robust against false positives (see e.g., Foote et al 2023).

Adopting a Bayesian approach, we tested the ability to infer the presence of life as a function of its (currently unknown) prior probability, using the Archean Earth-like atmospheric data we generated as a case study. To do so, we calculated the likelihood of life as the best explanation for a given observational value, *P(life|O)*, of gas abundance or network metric as follows:

$$P(life|O) = \left(\frac{P(O|life) \times P(life))}{P(O)}\right) \tag{1}$$

Here *P(life)* is the prior probability for life on an exoplanet, *P(O|life)* is the posterior probability to observe a given signal (e.g., a particular value of methane gas abundance or a network metric value) produced by life; and *P(O)* allows normalization to the total likelihood of that particular observation being made (e.g., the probability of any exoplanet (inhabited or not) sampled from a random distribution to have a specific value of methane gas abundance or network feature). The

target of calculating *P(life|O)* therefore captures how likely we can consider the hypothesis of life as the correct explanation, given observation, *O*.

Using probability distributions for each observation value, derived from the datasets (see **Supplement Figures S6-S12**) and treating *P(life)* as a free parameter, we evaluated *P(life|O)* over a range of *P(life)* values spanning four orders of magnitude, with *P(life)* = 0.001, 0.01, 0.1, 0.5, and 0.9, see **Figure 3**. We compared our models of known abiotic production of $CH_4$ and two anomalous production mechanisms to the case of a known biological mechanism (i.e., the explicit biotic flux model scenario). In other words, we assess the probability of life detection based on pairs of possible scenarios: abiotic vs biotic, anomalous surface flux vs biotic, and anomalous steady-state vs biotic. Our goal is to address the question: How easy is it to distinguish one case from the other, based on observations that are either a specific abundance, a global non-network-based metric, or a particular network metric? Note, the anomalous surface flux scenario could be either associated with life or an abiotic process: our assumptions here are only that we do not know the source of a methane flux, but the flux is the same rate as from known methanogenesis reactions. In addition, *P(life)* here implies *P(life using Earth-life methanogenesis)*. The results show that for a large fraction of the metric-space, we need high prior confidence in the existence of life (large *P(life)*) to reach a high value of *P(life|O)*. Mean degree <*k*> and average shortest path length <*l*> are the network metrics least sensitive to the observational data paired with the biotic model (abiotic or anomalous), but they still exhibit a significant sensitivity to *P(life)* in the anomalous cases towards large values. However, some regions are associated with a high probability of life detection while being largely insensitive to *P(life)*, such as a high methane abundance, a very high mean degree <*k*> or very high average shortest path length <*l*> in the abiotic vs biotic comparison, or a very low methane abundance or medium values of edge betweenness centrality in the anomalous steady-state vs biotic comparison. Any of these cases might lead to high-confidence in life detection, independent of our prior knowledge of how likely life is.

Even more conclusively, some network metrics appear to be very strong indicators of the *absence* of life-as-we-know-it, such as mean degree <*k*> ( for values $<10^{13.7}$), and average shortest path length <*l*> ( for values $<10^{12.7}$) – see grey regions of **Figure 3**. Thus, low values of these network metrics can be treated as an "*anti-biosignature"* for known metabolism: in the case of sufficiently low values, we can confidently rule out the hypothesis of a known mechanism of biologically driven flux when compared to abiotic or anomalous explanation. The same is true for high values of clustering coefficient <*C*> (for values $>10^{4.2}$), which also acts as an anti-biosignature. This is especially useful for ruling out the possibility of Earth-like methanogenesis on an exoplanet, where a focus only on $CH_4$ abundance alone would yield ambiguity depending on the value of *P(life*) used (see **Figure 3**, top panel), corresponding to cases with $CH_4$ abundance between $10^1$ and $10^3$ ppm. That is, using a complex systems-based approach allows ruling out hypotheses we might otherwise be unable to.

Taken together, these results encourage the combination of statistical approaches of network analysis and Bayesian inference to determine whether some observational data is driven by the presence of alien life. Our Bayesian analyses were conducted pairwise on the datasets with the goal to test an explicitly biotic scenario against different well-defined hypotheses. The pairwise comparison enables clearer results and interpretation for this exploratory study, however future work should consider more of these scenarios and treat all the variables as possible observations, in order to estimate *P(life|any source of observation)* across all viable hypotheses simultaneously. We note that in our analyses we treat network measures as observables, on equal footing with methane gas abundance, whereas only the latter is directly inferable at present from spectral data. The global species variance metric also requires knowledge of all the species and abundances in the system and performed worse than methane abundance and any other network metric, showing that it is not enough to focus on any system-level property. The network measures, by contrast, are demonstrated to capture those system-level features that might be more directly relevant to life detection. We discuss in the Discussion section opportunities for future work aimed at directly inferring network topology from spectral data, which would put network measures on equal observational footing with gas abundances.

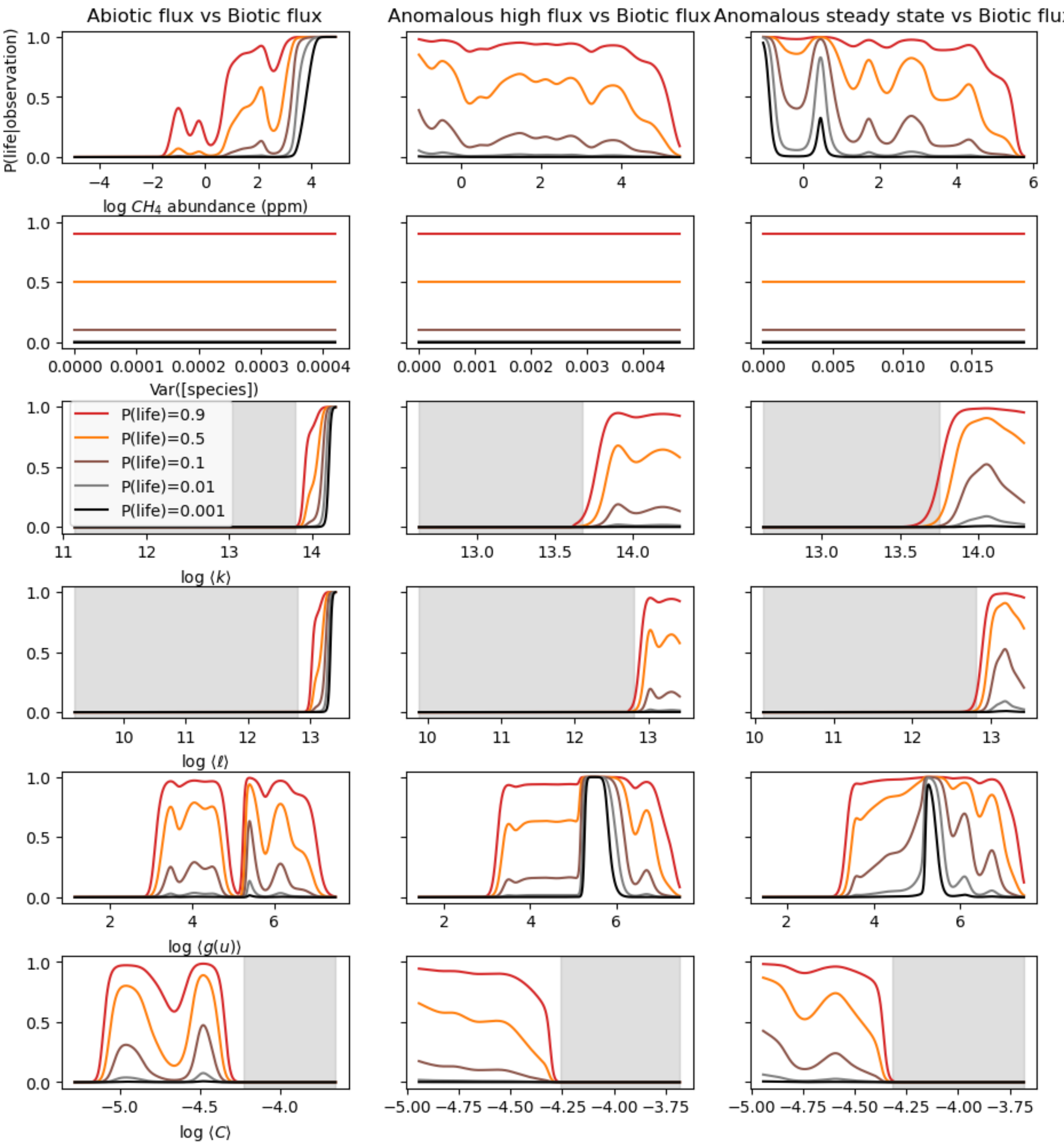

***Figure 3.*** *Bayesian analysis of the posterior likelihood of life, P(life|observation), assuming an observed value for network metrics, variance of species abundances, or methane gas abundance, probed at 0.2 mbar. Given that the probability for life to exist on an exoplanet is unknown, we assess the posterior likelihood over a range of prior probabilities for life, P(life). Grey regions highlight metric values that are associated with a zero probability of life and are therefore anti-biosignatures for the sets of models compared herein. In general, the subset of network metrics studied here can provide greater constraints on the probability of life given a set of observations than* $CH_4$ *abundance alone, allowing distinguishing cases where Earth-like methanogenesis can be ruled out even when* $CH_4$ *abundance alone is not conclusive.*

## Extension and Application to Modern Earth-like Atmospheres with and without Technologically Produced Gases

Having demonstrated the utility of our network-based approach for analyzing Archean Earth-like atmospheres, we extend our analyses to examine a "Modern Earth-like" scenario. Our goal is to test whether our methodology can also distinguish atmospheres influenced by technological activity from those without such influence. Technosignature gases are biosignatures (Wright et al. 2022) produced by technology. To investigate a technosignature scenario, we compare a biosphere with and without industrially produced CFCs. Specifically, we compare a case when the atmosphere is consistent with the presence of a modern Earth-like biosphere (but not a technosphere, e.g. Earth several centuries ago) to that of a biosphere which evolved to produce industrial gases (our modern technosphere). While we focus on an industrially produced gas, we note that our goal for developing network-based methods is specifically because they should be generalizable to scenarios of unknown technological origin – i.e., gases that may be technosignatures but are not produced by industrial processes as we would recognize them on Earth. We define two scenarios: one case corresponding to an atmospheric scenario with CFC-12, and one case without CFC-12, see **Figure 4**. The two scenarios are as follows:

1. **Modern Earth.** The simulated atmospheres follow the modern Earth analogue dataset developed by the NASA Frontiers Development Lab, which provided a dataset of over 120,000 modern-Earth-like atmospheric simulations, with varying amounts of $CH_4$, $CO_2$, $H_2$, and $O_2$. From this dataset, we randomly selected 5,000 models as our ensemble of Modern Earth scenarios. These represent an Earth-like planet, consistent with a biosphere but without the presence of industrial emissions.
2. **Modern Earth with Industrial Production of CFCs.** The same 5,000 Modern Earth scenarios were re-run, incorporating the presence of $CCl_2F_2$ in the atmosphere, also known as CFC-12 or Freon™. CFC concentration was of the form $x \times 10^{-7}$, where x was drawn from a Gaussian distribution centered on a mean of 0.5, with a standard deviation of 0.5, leading to concentrations ranging from $10^{-11}$ to $10^{-6}$ ppm. This corresponds to a biosphere with the presence of an unequivocally industrially produced technosignature gas.

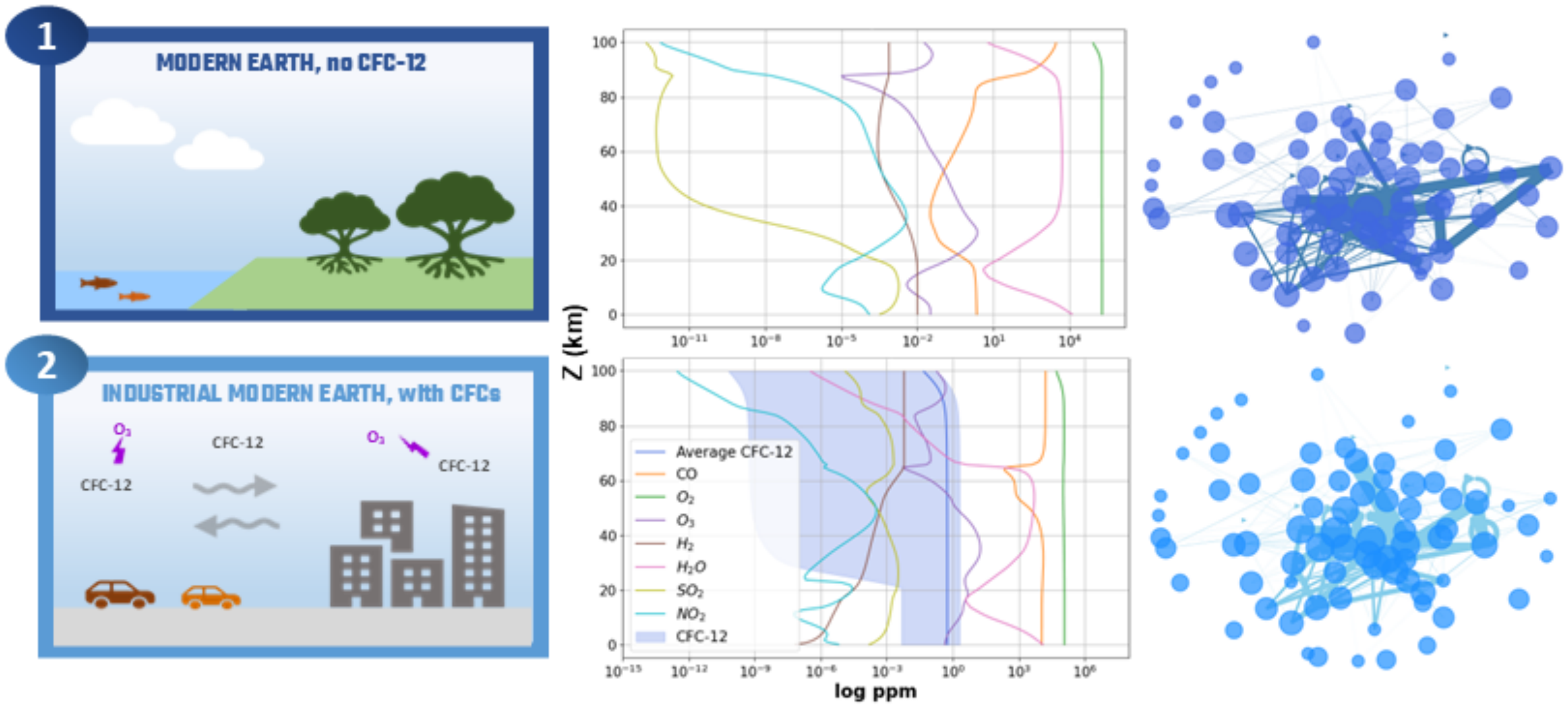

***Figure 4:*** *Conceptual representations, abundance profiles, and network visualizations for each category of Modern Earth atmospheric model used in this study. The first column shows the scenario, the second column shows the vertical profiles of the atmospheric species abundances corresponding to the initial conditions for each scenario and the third a network representation of the chemistry in the atmosphere. From top to bottom: 1)* ***Baseline modern Earth model****. This model includes no industrial CFC-12 emissions. 2)* ***Modern Earth modeled with emissions of CFC-12****. A technosphere is modeled by the addition of a technological source for chloride radicals that can catalyze the destruction of $O_3$.*

In this study, CFC-12 is the candidate technosignature gas of interest, analogous to $CH_4$ as the biosignature gas of interest in the Archean Earth studies. Notably, the abundance of the technosignature gas of interest, CFC-12, cannot be compared between the two models because there is no CFC-12 in the model without industrial activity. Therefore, the only measures to compare between the two cases are system-level measures or agnostic network measures that do not refer to specific molecular species.

The modern Earth analogue model yields a network with 70 nodes and 560 edges for the baseline case, and 75 nodes and 574 edges with the introduction of the technosignature gas CFC-12. As was the case for methane production, each network measure can distinguish between the distributions for the different scenarios, see **Figure 5**. Mean degree, <k>, yielded the largest dissimilarity between the biosphere and technosphere atmospheres (top row **Figure 5**), consistent with the interpretation of mean degree tracking the presence of highly reactive species (see **Table 1**). This analytical approach could potentially be applied to exoplanets harboring a technosphere, which produces gases that are not recognizable as being the same as industrially produced gases on Earth.

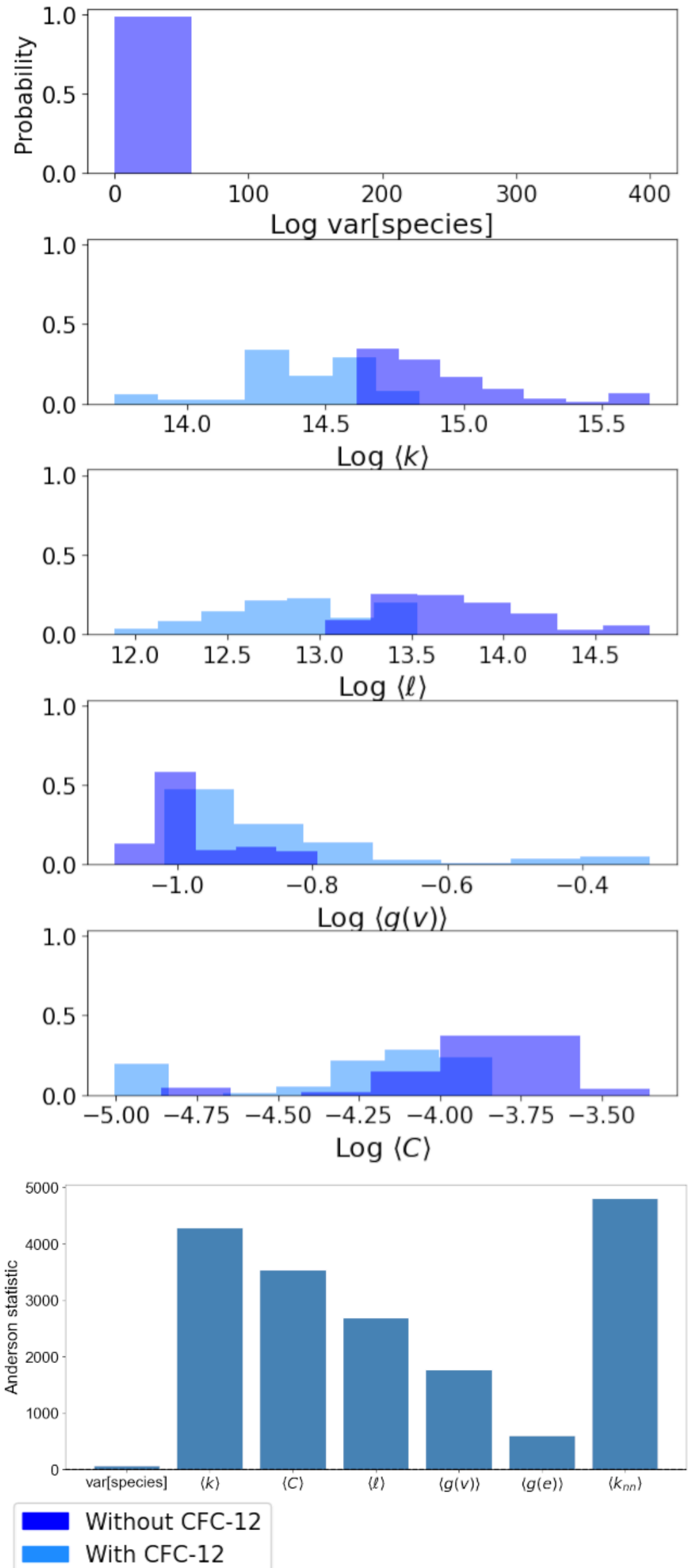

***Figure 5***. *Top: Distributions of the probability (y-axis) for a simulated atmosphere to yield the given species abundance variance or network metric value (x-axis) for datasets sampled from modern Earth atmospheres with and without the presence of CFC-12. Bottom: Anderson scores quantifying the reliability of each network metric to distinguish between atmospheres with and without CFC-12, where the threshold for distinguishability is at 6.546 (0.1% significance level).*

# Discussion

Our analysis of both Archean Earth-like and Modern Earth-like atmospheric datasets demonstrates how complex systems science approaches can be used as an effective tool for distinguishing features of atmospheres driven by known biological or technological processes, when compared to those driven by known abiotic processes, and those that are anomalous (not confirmed as a known abiotic, biological or technological process). Our results demonstrate the utility of such approaches for ruling out concrete hypotheses that are not consistent with the data. In some cases, this may include ruling out the hypothesis that alien life on an observed exoplanet uses the same metabolism or industrial processes as on Earth.

Across both the Archean Earth-like and modern Earth-like atmospheric datasets, network structure exhibited quantitative changes to the introduction of biogenic or technogenic reaction pathways. In the case of the Archean Earth group, the metrics that were most useful in distinguishing between the biotic and abiotic models were those that were sensitive to changes across the entire network, such as clustering coefficient and node betweenness centrality, which both capture features of how individual molecular species can cause changes in many other species' abundances and reaction rates due to nonlinearities in reaction pathways and their interactions. In the models presented here, this manifests in how the species most affected by the introduction of the biological methanogenesis – methane – is not, in of itself, particularly reactive. Instead, the change in network topology is due to the addition of the methanogenesis reaction itself, which provides a 'shorter' and more efficient route for the reduction of $CO_2$ to $CH_4$ than what is possible in the absence of biology. Thus, this additional pathway can dominate the behavior of the network by tilting the balance of the atmosphere away from $CO_2$ and towards $CH_4$.

In contrast, the metrics that were most useful in distinguishing between modern Earth atmospheres with and without CFC-12 emissions were those that can be sensitive to changes in individual nodes, such as degree. Indeed, from one network to the other, the CFC-12 node gains two edges while the Cl radical it generates gains 12. In other words, while the throughput of reactions involving CFC-12 is not very large, the resulting chloride radical is extremely reactive and has a disproportionately higher impact on the chemistry of other species in the atmosphere with respect to its abundance, leading to a higher degree in the graphical representation of the atmosphere and skewing the global statistic of mean degree.

Application of Bayesian analyses demonstrates that for the hypotheses compared in this study, network metrics were not necessarily able to provide a significantly stronger confidence in the detection of alien life compared to $CH_4$ abundance. However, they were able to provide much stronger confidence in *not* detecting methanogenesis as it exists on Earth. This is arguably equally interesting considering that the breadth of abiotic scenarios on exoplanets is poorly constrained and that better priors on the absence of life will build up to a much greater confidence in life detection claims further on. Our work shows how network metrics may be valuable as "anti-biosignatures" (Schwietermann et al 2019, Wogan and Catling 2019), with, for example *log*<*k*> or *log* <*l*> smaller than 13 being strongly indicative of the absence of methanogenesis as found in life-as-we-know-it. We stress that in the context of our study, an anti-biosignature constitutes evidence for the absence of a global biosphere impacting the atmospheric chemistry (specifically through methanogenesis, a subcase of "life-as-we-know-it"), rather than necessarily a complete absence of life. Finally, the overall bad performance of the metric of species abundances compared to network metrics shows the usefulness of this approach in capturing physical properties of the atmosphere that non-network measures operating at the system-level cannot necessarily capture.

Taken together, a unified approach to the study of a biosignature gas (methane) and a technosignature gas (CFC-12) suggests that network analyses can pick up system-level changes that arise due to the presence of life producing either simple gases that drive global changes in atmospheric network structure, or more complex gases that drive changes in atmospheric network structure. We therefore conclude that introducing the study of chemical reaction network topology in atmospheric characterization pipelines could prove useful as an agnostic method for assessing the presence of life on a planet based on its chemical by-products. This is particularly true in cases where the underlying biosphere or technosphere may be unknown or poorly characterized, because we can rule out Earth-like biology or technology and distinguish it from cryptic anomalous signals that might otherwise be difficult to distinguish from known abiotic or biological sources.

***The importance of weighted networks***

While previous studies have proposed network analysis as a possible orthogonal method for assessing the presence or absence of life via atmospheric chemistry (e.g., Wong et al. 2023), our study provides new insights by quantifiably examining a weighted network built from a large ensemble of chemically self-consistent simulations. Importantly, weighted networks provide critical advantages over unweighted networks because the latter comprise a list of abiotic chemical reactions, which are empirically derived in the laboratory or via ab initio methods but are not necessarily specific to a given planetary context. These "unweighted" networks are derived by assessing which minimum set of reactions are needed to explain some aspect of planetary atmospheric chemistry to some uncertainty level. However, in exoplanet modeling much work is based on the idea of building universal photochemical models that contains many possible gas phase (photo)chemical reactions and are tuned to simulate specific atmospheres, given relevant boundary conditions that constrain rates to fit specific planetary contexts. The topology of an unweighted network derived from such a "universal" network containing all atmospheric reactions could allow inference about life if one considers a threshold to the weights that are specific to an individual planet. Our view, supported by the analyses we presented herein, is that weighted networks will likely allow for deeper insights by incorporating the relative significance of reactions, which are sensitive to planetary conditions and stellar environment, enabling identification of potential biogenic inputs and outputs within the atmosphere.

***Observability considerations***

We hope to investigate further the utility of this approach by testing it on a wider variety of atmospheres and biospheres, such as planets with an abiotic $O_2$ flux (Kleinböhl et al. 2018), or the complex chemistry of Titan-like worlds (Willacy, Allen, and Yung 2016). A next step is to quantitatively link these modeling studies to current and future observational prospects. The most straight-forward way is to determine whether the network topology of an atmosphere can be directly inferred from exoplanet spectra by building forward models that account for network features, and subsequently using these to build new pipelines for atmospheric spectral retrievals (Madhusudhan, 2018; Barstow & Heng, 2020). For example, it should be possible to compare networks obtained for a few observed exoplanets with known theoretical network "templates" at first, e.g. by fitting two separated (fully defined) networks and comparing their Bayesian evidence. Eventually, the goal would be to build a more holistic understanding of how topological properties are influenced by different (bio)chemical processes, removing the need to work from fully defined templates. Overall, if network properties can accurately be included in atmospheric spectral retrievals, then their associated spectral signatures – which may not be directly dependent on the observed abundances of biologically relevant gases – would be a powerful addition to the toolbox of exoplanet-focused astrobiologists.

If direct inference of network structure from exoplanet spectra is not possible, then the next challenge is to determine the minimum network size (both in terms of species identified, and uncertainties in abundance) required to meaningfully characterize an exoplanet's atmosphere. This goes together with determining the extent to which network properties are robust against uncertainties in the environmental conditions as output by the retrieval. We have shown in this study that some network metric values provide more confidence towards detecting the presence or absence of life than a more constrained abundance estimation of the biologically relevant gas. In addition, developing new frameworks for the analyses and interpretation of exoplanet atmospheric data becomes especially important when the atmospheric gas abundances of interest are too low to be detectable. In the case of methane, in addition to being poorly reactive, it is also weakly soluble, with solubility being a clear practical factor in assessing which biogenic gases can potentially accumulate in the atmosphere in the first place (Seager et al. 2012). A 10 ppm-noise floor is typically assumed in the case of JWST's Near-Infrared Spectrometer or Mid-Infrared Instrument (approximately 2 dex broad abundance characterization), which is too large to detect $CH_4$ in a significant part of the Archean Earth simulations of this study, and too large to detect CFCs in all the Modern Earth simulations given that the concentrations never reach 1 ppm. Including network analyses could significantly reduce the number of co-added transit time needed to infer their presence in the atmosphere (currently estimated at ~100 h in the case of CFCs according to Haqq-Misra, Kopparapu, et al. 2022). The size of population studies is also an important factor to take into consideration for comparison of different network properties across a variety of atmospheres. JWST will provide insights on a few major species (mostly $CO_2$, $H_2O$ and $CH_4$) for approximately 20 planets, but the constraints will be inhomogeneous given that various instruments are being used, with different resolution and wavelength coverage. ARIEL will enable a more consistent sampling and characterization of hundreds of exoplanet atmospheres (divided into 3 Tiers) but will likely be limited to larger and hotter planets than the terrestrial ones investigated in this study (probably not smaller than the Sub-Neptune/Super-Earth regime). While providing much better abundance constraints (approximately 1 dex) for species like $H_2O$, $CH_4$, CO, $O_2$ and $O_3$. The Habitable Worlds Observatory (currently discussed for late 2040) will focus on studying in detail only a few (<25) terrestrial exoplanets, thus already warranting a good statistical understanding of how complex system approaches fit in the broader atmospheric landscape of exoplanets before making confident life detection claims. These observational surveys rely on knowledge of chemical kinetics and molecular cross-sections for spectral retrievals that are still incomplete and/or uncertain. While this poses challenges for network analyses (as it does for any other approach), forward modeling of network retrievals could allow us to determine the most important kinetic reactions to refine with further laboratory experiments or analytical calculations, in order to increase our confidence in exoplanet data interpretation. Overall, it is unclear at this stage what planetary properties are essential to reliably construct the chemical reaction network of an exoplanet atmosphere, and future work should focus on identifying the best set of observables for network-based retrievals.

In summary, not only would a network theoretic framework increase the overall confidence in a best-fit planetary scenario for any observed atmosphere, but it also offers a more consistent and agnostic approach for the detection or ruling-out of planetary biospheres influencing atmospheric chemistry. It also opens new considerations for future observation campaigns, such as determining the best suite of spectrally active species beyond biogenic gases themselves, maximum uncertainties in the temperature structure of the atmosphere, required spectral resolution and wavelength range, and optimal combination of observations/observatories, etc.

Finally, beyond the implications for biosignature detection, the influence of biology on atmospheric reaction network structure may provide a window into the physics of life itself. If there is, as has been speculated, a 'universal biology' dictated by the physical constraints of the universe (Mariscal and Fleming 2018), one manifestation of it may be in its network topology (Solé et al. 2002). Given how tightly coupled the biosphere is to the atmosphere on Earth, it is not unreasonable for the biosphere to influence the chemical reaction network topology of the atmosphere. For instance, one might model planetary evolution as a multilayer network, where each layer represents the chemistry in the geosphere, biosphere, or atmosphere (and technosphere for planets where technology has evolved (Frank et al 2022)). The impact of industrial emissions on atmospheric reaction network topology is harder to explain in this manner, though similarities in the topology between biological and technological networks are not unprecedented (Solé, Valverde, and Rodriguez-Caso 2011). In any case, further investigation into atmospheric reaction networks is warranted, with interesting implications in a variety of fields of exoplanet science and astrobiology.

# Acknowledgements

This work was supported by award # 80NSSC21K1402 awarded by the National Aeronautics and Space Administration. The authors would like to thank Shawn Domagal-Goldman, Giada Arney, and Sonny Harman for their assistance with implementing `atmos`. We thank Hyunju Kim and Cole Mathis for valuable feedback throughout various stages of the research and manuscript preparation, and Quentin Changeat and Harrison Smith for reviewing the manuscript before submission. We also thank the reviewers for valuable feedback and discussion about relevant statistical measures and applications of network theory to planetary atmospheres.

# Author Contributions

TF identified, implemented and ran atmospheric models, surveyed the literature, developed the network topology analysis pipeline, produced data figures, calculated statistical metrics and predictiveness scores and wrote the initial draft. EJ provided technical details for observatories and exoplanet models, helped with the analysis, produced conceptual figures and wrote the paper. SIW conceived of the study, supervised the project and revised the manuscript, with the support of EJ and TF.

# Appendix

## Generating Ensembles of Simulated Exoplanet Atmospheres with `atmos`

To test whether the topology of chemical reaction networks can be useful for exoplanet life detection, we simulated two sets of terrestrial atmospheres, one modeled after the Archean Earth with methane as a candidate biosignature gas, the other a modern Earth analogue to examine the effect of chlorofluorocarbon emissions as a technosignature gas. Simulations were run over a wide range of initial conditions, using the `atmos` modeling package (Arney et al. 2016). `atmos` is a combined photochemical and climate model, however the work in the present manuscript only makes use of its photochemical aspect. Atmospheric photochemistry is modeled via a stack of 200 plane-parallel layers from the surface to an altitude of 100 km, with 0.5 km layer spacing. The mixing ratio of species in the atmosphere is found by solving flux and mass continuity equations in each layer simultaneously using a reverse-Euler method, providing exact solutions at steady state. Vertical transport is represented via molecular and eddy diffusion and can be modified by user-set boundary conditions.

The photochemical model is considered converged when the redox chemistry is conserved and a re-run of the model using last run's output as initial conditions converges quickly (*i.e.*, <50 time steps, as opposed to the hundreds of time steps normally required). The model sets were run until they converged on a steady state solution for the atmosphere that would be stable over long time periods under constant inputs to the atmosphere. To construct as large a sample size of models as possible, we implemented a long cut-off of a maximum simulation time of $5\times 10^{10}$ simulated years, though most simulated atmospheres converged before reaching that point. Simulations that did not converge were not included in the dataset used for the network analyses.

Regarding parameters specific to `atmos` and for the sake of future reproducibility, we note that the lightning module was turned on (with the values for PRONO, NO production from lightning set to 1e9). Updated cross-sections from Ranjan et al. (2020) were not used and future studies would benefit from correcting the relevant reactions highlighted in this study. To make this study relevant to the large population of small rocky planets orbiting red dwarfs, this work uses the spectrum of the star GJ581 from the MUSCLES Treasury Survey (France et al. 2016) for both the Archean Earth and Modern Earth datasets. `atmos` then readjusts the semi-major axis of the planet such that it is subject to the same stellar flux as the Earth around the Sun (through a parameter called Solar Flux Scaling, or FSCALE).

## *Archean Earth-like Worlds with and without Biologically Produced $CH_4$*

We simulated 20,000 Archean Earth-like atmospheres with 200 vertical layers, using only the photochemical component of `atmos` and a fixed temperature-pressure structure (see **Figure S1**) to converge on consistent solutions for the chemical profiles (including aerosols). For the temperature-pressure profile, **Figure S1**, we used `atmos` ' default Archean Earth template (Arney et al 2015). To simulate many atmospheres, we used the `PyAtmos` Python wrapper to run batches of `atmos` models in parallel (Chopra et al. 2023). Each atmosphere had differing values for $CH_4$ fluxes but were otherwise identical. **Table S1** lists the boundary conditions and deposition velocities for all the gases present in the simulations.

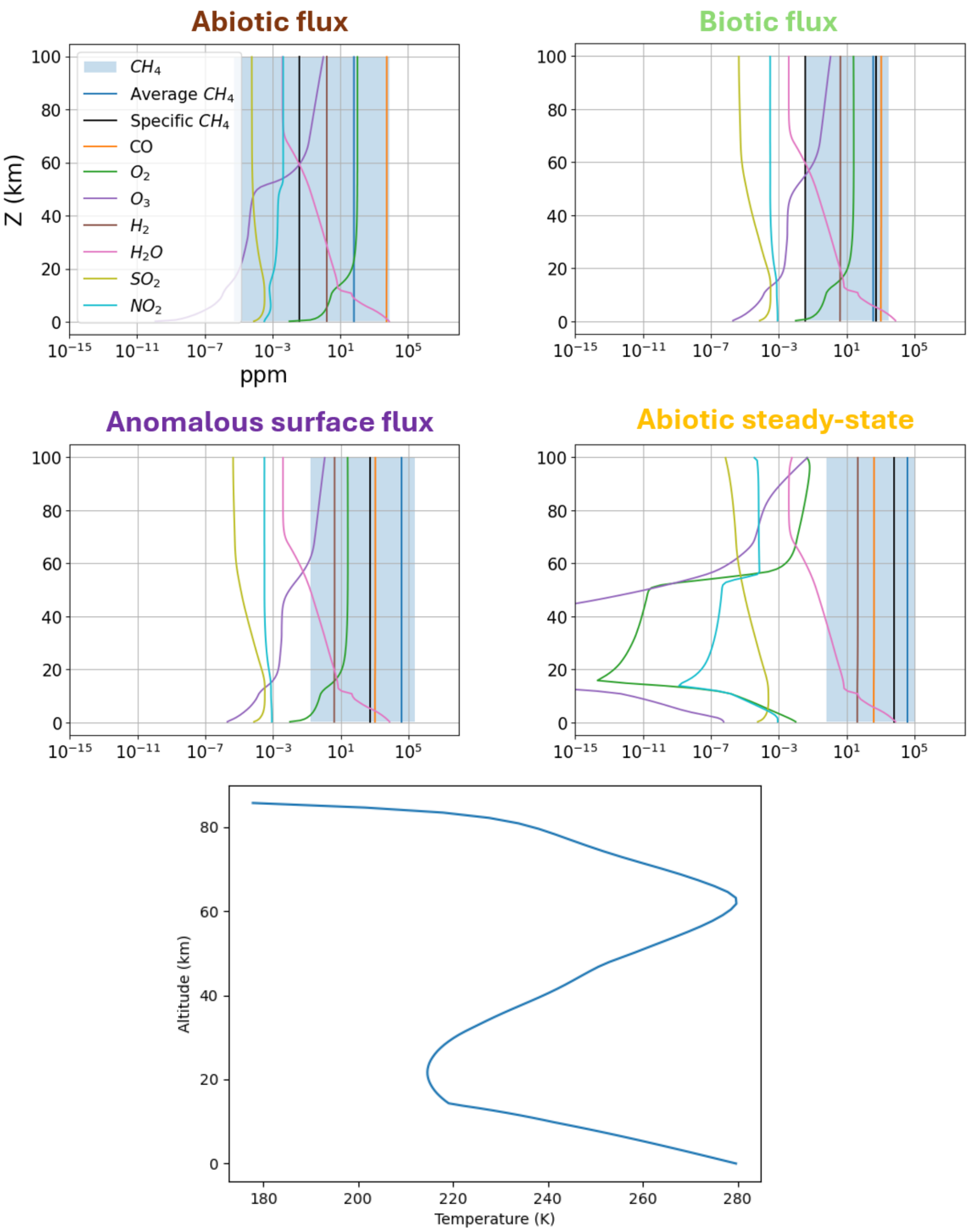

***Figure S1.*** *Top: Abundance of important atmospheric species as a function of scale height for each modeled scenario. Bottom: The temperature profile as a function of scale height. All the Archean Earth model scenarios are characterized by the same temperature-pressure profile, from Arney et al. (2016) – here, output of the converged* `atmos` *simulations.*

| Species | Constant deposition velocity | Constant upward flux |
|---|---|---|
| O | 1.00E+00 | |
| H | 1.00E+00 | |
| OH | 1.00E+00 | |
| $HO_2$ | 1.00E+00 | |
| $H_2O_2$ | 2.00E-01 | |
| CO | 1.20E-04 | |

| HCO | 1.00E+00 | |
|---|---|---|
| $H_2CO$ | 2.00E-01 | |
| $CH_3$ | 1.00E+00 | |
| NO | 3.00E-04 | |
| $NO_2$ | 3.00E-03 | |
| HNO | 1.00E+00 | |
| $O_3$ | 7.00E-02 | |
| $HNO_3$ | 2.00E-01 | |
| HSO | 1.00E+00 | |
| $H_2SO_4$ | 1.00E+00 | |
| $SO_4AER$ | 1.00E-02 | |
| $S_8AER$ | 1.00E-02 | |
| HCAER | 1.00E-02 | |
| HCAER2 | 1.00E-02 | |
| $O_2$ | 1.00E-08 | |
| $CH_4$ | | Varies |
| $CO_2$ | | 2.00E-02 |
| $H_2$ | 2.40E-04 | 1.00E+10 |
| $H_2S$ | 2.0E-02 | 3.500E+08 |

***Table S1.*** *Lower boundary conditions for the Archean Earth model. The species marked with "AER" are atmospheric aerosols.*

Due to limitations in the `atmos` model implementation only two reactants are allowed per reaction, without stoichiometric coefficients. The methanogenesis reaction $CO_2 + 4H_2 \rightarrow CH_4 + 2H_2O$ in the biotic surface flux model was therefore included in the CRN as a three-step reaction where the kinetics of methanogenesis for anaerobic microbes are modeled by an Arrhenius equation, using rate constant K (in kJ/C mol hour):

$$K = 3.3 * e^{\left[\left(\frac{-6.94\times10^{4}}{R}\right)-\left(\frac{1}{T}-\frac{1}{298}\right)\right]} \tag{1}$$

(Tijhuis, Van Loosdrecht, and Heijnen 1993; Seager, Bains, and Hu 2013b).

To accommodate this modeling limitation, our implementation involves several fictitious and short-lived intermediate species, that all have the rate constant K:

1. $H_2 + H_2 \rightarrow$ '$H_4$'

2. $H_4 + CO_2 \rightarrow$ '$H_4CO_2$'

3. '$H_4$' + '$H_4CO_2$' $\rightarrow CH_4 + 2\ H_2O$

These fictitious intermediates are short-lived species, which are modeled separately in `atmos` from long-lived species, and are excluded from the model's Jacobian. Though these species are included and therefore affect the network topology, they only participate in six edges, and their short lifespan leads to low edge weight values ($<10^{-73}$). As a result, their impact on the weighted topological metrics is minimal, see **Figures S2 and S3**. This reaction is the only one requiring modification with fictitious species in our models. Note that $CO_2$ and $H_2$ remain fixed so we ignore their dynamic effect on the methanogenesis rate, whose value is sampled from a Gaussian distribution centered around $10^{11}$ molecules/cm$^2$ and a range between $10^3$ molecules/cm$^2$ and $10^{13}$ molecules/cm$^2$ (cutoff value), which is consistent with methanogenesis on Earth during the Archean (Arney et al. 2016).

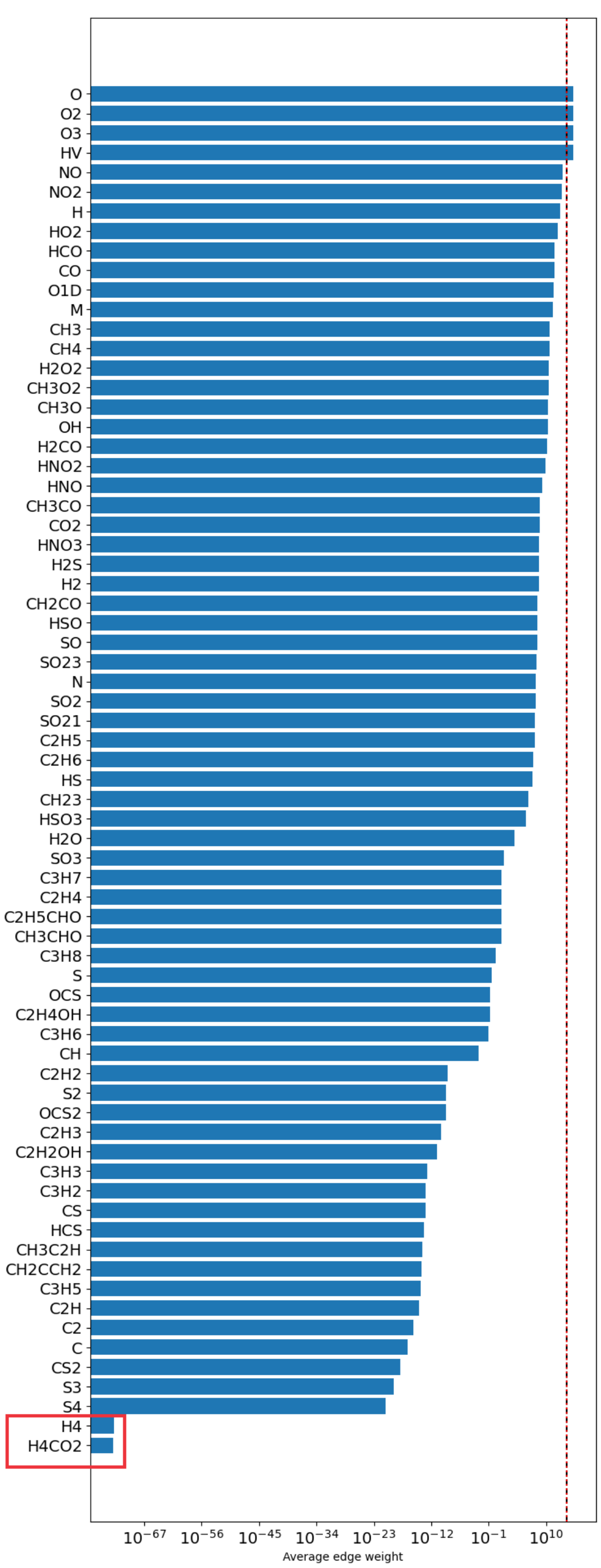

***Figure S2***. *Ranked edge weights of each species in the biotic model, calculated by taking the average of edge weights for each node, and then averaging that value over the entire ensemble. The black dashed line is the average of these values for all nodes when the fictitious species are*

*present, and the red dashed line is the average of these values for all nodes without the fictitious species. The inclusion of the fictitious species $H_4$ and $H_4CO_2$ (outlined by the red box) minimally affects average network statistics computed from edge weights.*

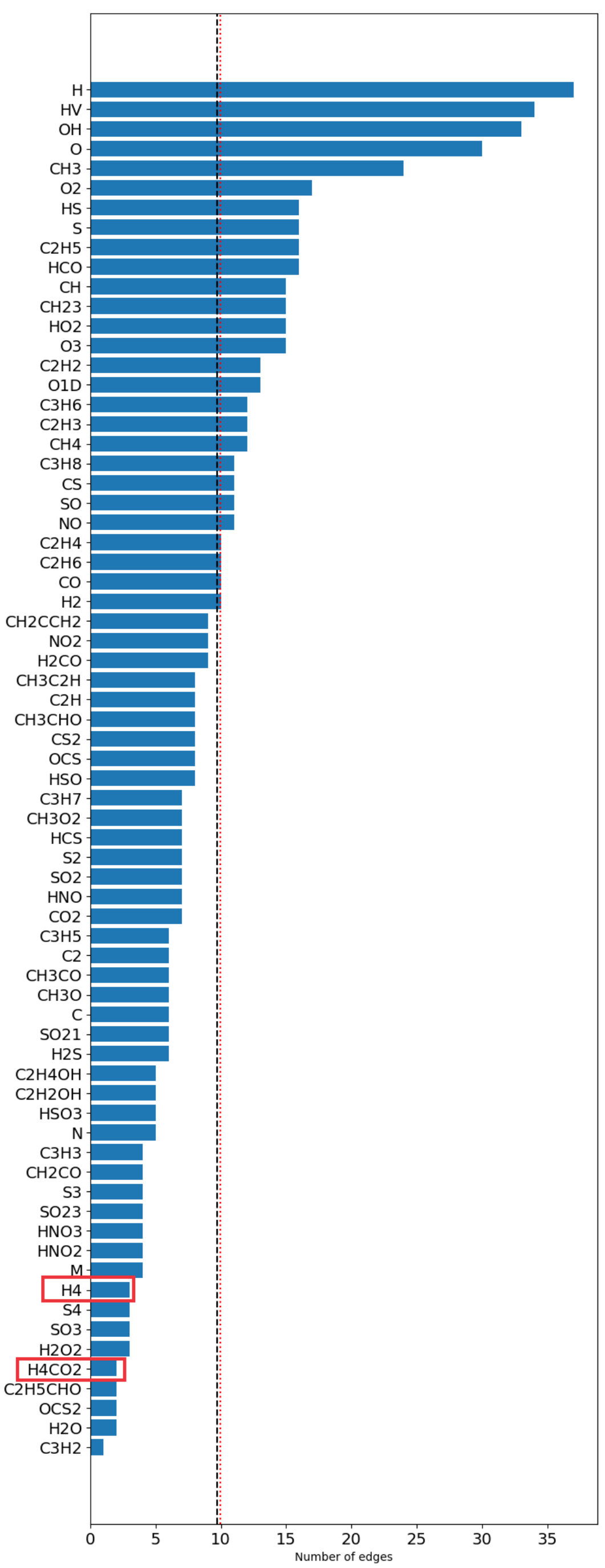

***Figure S3**. Number of edges for each species. The black dashed line is the average of these values for all nodes when the fictitious species are present, and the red dashed line is the average of these*

*values for all nodes without the fictitious species. $H_4$ and $H_4CO_2$ (outlined by the red boxes) both have comparatively few edges, and their presence or absences do not effect average values.*

## Constructing and Measuring Atmospheric Reaction Networks

From each of the modeled atmospheres, we created a network representation of the chemical reactions in the atmosphere and measured topological features of the created graphical representations of the atmosphere to characterize the structure of these networks. For models that did converge to a steady state (ranging between 41% and 95% of models run, depending on the model category and abundance of $CH_4$, as high abundances can cause the model to become sufficiently 'stiff' and fail to converge in a timely manner), the abundances and reaction rates were extracted. To construct networks from these modeled atmospheric data, we represented each species present in an atmosphere mathematically as a *node* within a unipartite network, with a node connected to another node by an undirected link if they co-participate in the same reaction as a reaction-product pair; this connection, or *edge*, is weighted by the reaction rate, see **Figure S4**.

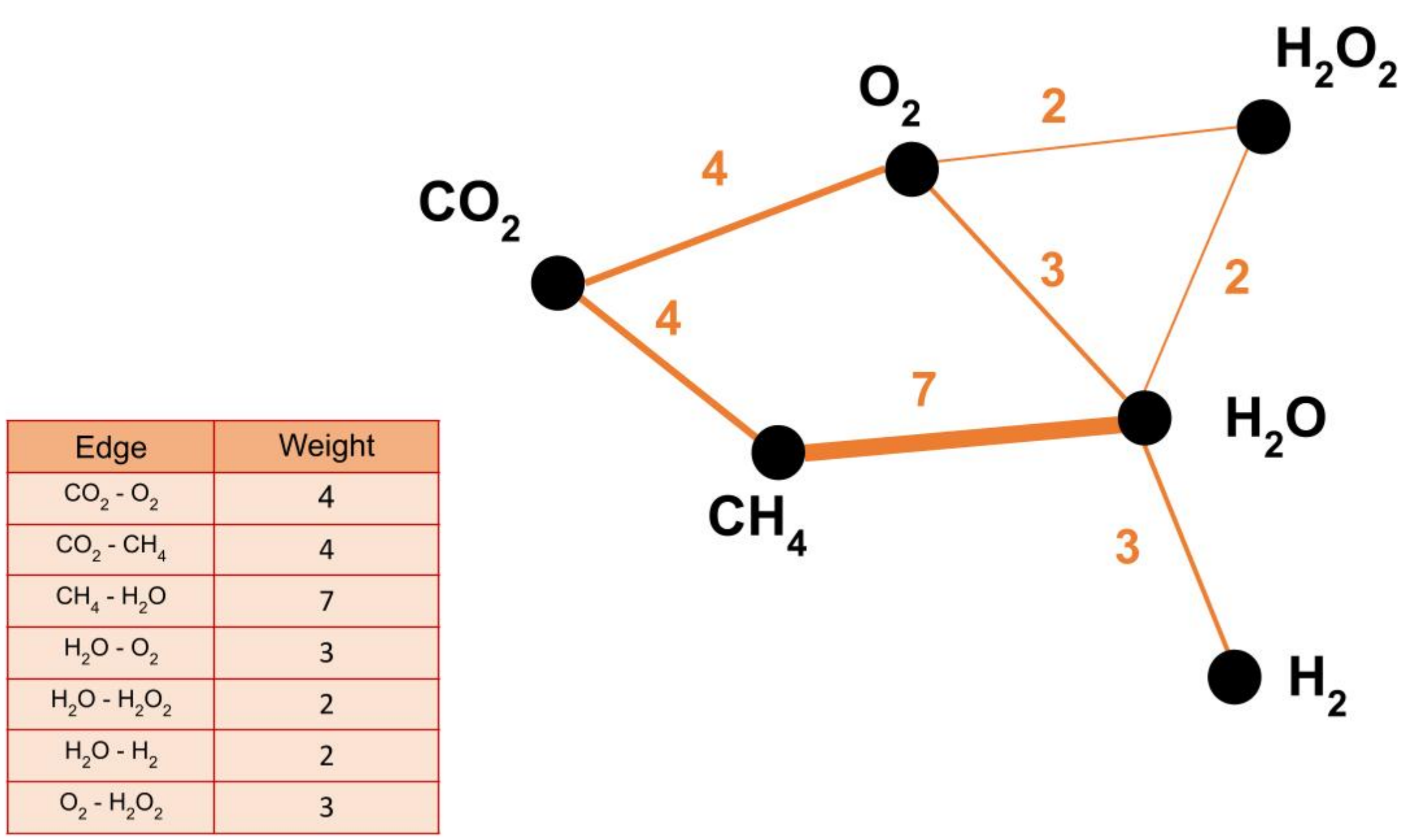

| Edge | Weight |
|---|---|
| $CO_2$ - $O_2$ | 4 |
| $CO_2$ - $CH_4$ | 4 |
| $CH_4$ - $H_2O$ | 7 |
| $H_2O$ - $O_2$ | 3 |
| $H_2O$ - $H_2O_2$ | 2 |
| $H_2O$ - $H_2$ | 2 |
| $O_2$ - $H_2O_2$ | 3 |

***Figure S4***: *A simplified example of a set of chemical reactions represented graphically as a network. The reactions included in this network are: $CH_4 + O_2 \rightarrow H_2O + CO_2$, $H_2 + O_2 \rightarrow H_2O$, and $H_2O_2 \rightarrow H_2 + O_2$. Each species present in the reaction list is represented by a node; reactant nodes are connected via edges to the nodes representing the products of the reactions they*

*participate in. Edge-weights correspond to the rates of the reactions represented by the edges, calculated as the reaction rate constant multiplied by the abundance of the reactants (edge weight values shown in this figure are for illustrative purposes only).*

To measure the topological parameters of the atmospheric CRNs, we fed the networks into a Python analysis pipeline, incorporating the tools from the NetworkX Python package (Hagberg, Swart, and S Chult 2008). We briefly summarize the network metrics used in this study in what follows.

***Degree*** is one of the most frequently used network measures to characterize the structure of a network (Jeong et al. 2000; Smith, Kim, and Walker 2021). For an unweighted network, the degree of node $i$, $d_i$ is simply the number of edges connecting the nodes to the rest of the network. Thus, in a chemical network, the degree quantifies the number of chemical species that a given species shares a reaction with. However, for a weighted network like the atmospheric network models we study here, the degree of node $i$, $k_i$ is the *sum of weights* of the edges connected to $i$:

$$k_i = \sum_{j \in V} a_{ij} w_{ij} \tag{2}$$

where $a_{i,j}$ is an element of an adjacency matrix of a given network and is equal to 1 when $i$ and $j$ are connected by an edge and 0 otherwise. This is the total sum of reaction rates for all reactions where a given chemical compound participates as a reactant or a product, which is equivalent to the *total flux* defined as the sum of total in-flux from and out-flux to all its neighbors in the atmospheric network. Note that the total flux defined here is different from the net flux defined as the in-flux subtracted by the out-flux. The average of the total flux over the set of all chemical compounds in the atmosphere can be quantified by the ***mean degree***, $\langle k \rangle$, defined as the average weight over all edges in a network:

$$\langle k \rangle = \frac{1}{N} \sum_{i \in V} k_i \tag{3}$$

where $N$ is the total number of nodes (chemical compounds) in the network.

The ***shortest path length*** from $i$ to $j$ for a weighted network is defined as

$$l(i,j) = \min_{\gamma(i,j)\in\Gamma(i,j)} \left[ \sum_{u,v\in\gamma(i,j)} w_{uv} \right] \tag{4}$$

where $\gamma(i,j)$ is a path from $i$ to $j$ and $\Gamma(i,j)$ is the set of all possible paths from $i$ to $j$ (Antoniou and Tsompa 2008). Note in a bipartite network the path would correspond to the more familiar definition of pathway as a linear sequence of reactions. However, in the unipartite representation we use here, the path should instead be thought of as the dependence amongst chemical species in a series of reactions where species $i$ and $j$ are participants in the end member reaction in each set). Hence, the shortest path length from compound $i$ to $j$ in the chemical reaction network represents the smallest amount of the sum of fluxes of chemical compounds along a dependency pathway connecting the two compounds. The **average shortest path length** over all pairs of nodes in the network can be written as

$$\langle l \rangle = \frac{1}{N(N-1)} \sum_{i,j\in V} l(i,j) \tag{5}$$

and this can be considered as the average minimal amount of total flux along a dependency pathway between every pair of compounds in the network.

***Clustering coefficient*** indicates the weighted density of edges between neighbors of a given node:

$$C_u = \frac{1}{d_u(d_u-1)} \sum_{vw} (\widehat{w}_{uv}\widehat{w}_{uw}\widehat{w}_{vw})^{\frac{1}{3}} \tag{6}$$

where $d_u$ is the number of edges connected to a node $u$, and $\widehat{w}_{uv} = \frac{w_{uv}}{\max(w)}$ is normalized by the maximum weight in the network (Onnela et al. 2005). $C_u$ is assigned to 0 for $d_u < 2$. A group of nodes with high clustering coefficient will be tightly knit.

***Average clustering coefficient*** (Barabasi 2016), is the average of clustering coefficients over all nodes and quantifies the tendency of interdependence between two chemical compounds sharing a neighbor compound:

$$\langle C \rangle = \frac{1}{N} \sum_{i \in V} C_i \tag{7}$$

***Average neighbor degree*** is the average weighted degree of the nodes connected to a given node (Barrat et al. 2004) and can be written as

$$k_{nn,i} = \frac{1}{k_i} \sum_{j \in N(i)} w_{ij} d_j \tag{8}$$

where $k_i$ is the degree of node $i$, $N(i)$ is the set of nodes connected to $i$, $w_{ij}$ is the weight of the edge that links nodes $i$ and $j$, and $d_j$ is the number of edges connected to node $j$ ($d_j$ is equivalent to $k_i$ in unweighted networks). The correlation between average neighbor degree and degree is useful for determining if a given network is *assortative* — such that high degree nodes tend to be connected to high degree ones and avoiding low degree ones — or if the opposite is true and the network is *disassortative* — such that high degree nodes tend to be connected to low degree nodes and not other high degree nodes. In atmospheric networks, it can identify whether chemical compounds with a high total flux tend to participate in the same reactions. In this paper, we computed the **mean value of average neighbor degree** $\langle k_{nn} \rangle$ over all nodes for each network. Note that in general, the mean value of the average neighbor degree alone cannot determine assortativity. However, in our atmospheric network models where a few nodes have a degree higher by several orders of magnitude than other nodes and dominate the total flux within the network, the high mean value can indicate the high degree nodes tend to be connected to each other, which indicates the network is assortative.

***Node betweenness centrality*** is defined as:

$$g(v) = \sum_{s,t \in v} \frac{\sigma(s,t|v)}{\sigma(s,t)} \tag{9}$$

where υ is the set of nodes, σ(s,t) is the number of shortest (s,t)-paths, and σ(s,t|υ) is the number of those paths passing through some node υ other than s, t, if s = t, σ(s,t) = 1, and if υ ∈ s,t, σ(s, t | υ) = 0. This measure can quantify the influence of a given compound on total flux along shortest

paths between any two compounds occurring within the network. Nodes with high betweenness centrality can sometimes be low degree, but essential to dynamics and function since they play a key structural role by connecting many otherwise disconnected or distant nodes. Hence it can be considered as global scale connectivity in contrast to the degree associated with local connectivity. We calculated the average node betweenness centrality, $\langle g(v) \rangle$, over each network. In atmospheric chemical reaction networks, the average node betweenness measures how often any given species is included in the shortest path between two other species in the network.

***Edge betweenness centrality***, is similarly defined to node betweenness centrality, except it represents the number of shortest paths that transverse through an edge, *e,* instead of a node:

$$g(e) = \sum_{s,t \in v} \frac{\sigma(s,t|e)}{\sigma(s,t)} \tag{10}$$

where υ is the set of nodes, $\sigma(s,t)$ is the number of shortest ($s,t$)-paths, and $\sigma(s,t|e)$ is the number of those paths passing through edge $e$. We calculated the average edge betweenness centrality, $\langle g(e) \rangle$, which captures the impact of edges within the network on large scale connectivity. This indicates how frequently any given reaction is part of the shortest path between any two species in the network.

# Quantifying Statistical Distinguishability Between Distributions

To test for statistical distinguishability between data generated from the different atmospheric scenarios, we used the Anderson-Darling (tail-weighted Cramer-von Mises) k-sample test, which is sensitive to a wide variety of differences between distributions (spread, tails, etc.). The Anderson-Darling test is a statistical test used to determine if a set of data was drawn from a given probability distribution. We made use of the non-parametric k-sample version of the test, which tests if two or more sets of data were drawn from the same distribution (Schloz and Stephens 1978), given as:

$$A_{kN}^2 = \frac{1}{N} \sum_{i=1}^{k} \frac{1}{N} \sum_{j=1}^{N-1} \frac{\left(NM_{i,j} - jn_i\right)^2}{j(N-j)} \tag{11}$$

where A is the value of the test statistic, N is the total number of observations in all samples, $Z_1 < \ldots < Z_N$ is the pooled ordered sample, and $M_{i,j}$ is the number of observations in the i-th sample that are not greater than $Z_j$.

Assuming a normal or exponential distribution, the critical value is calculated at a significance level of 0.1%. To ensure that comparisons between distributions were fair, we trimmed the data sets so that they all had the same number of networks as the smallest data set. This was done by removing the models with the highest $CH_4$ abundances, since these models were less likely to converge and are the least realistic. These trimmed data sets were also used in the Bayesian analysis described in a later section.

## Perturbative Analyses

To test the robustness of our results to missing data about the underlying atmospheric chemistry, we also ran a set of perturbative simulations where acetylene ($C_2H_2$) was removed. This species plays a key role in the formation of hazes: eliminating it removes a (poorly constrained) hydrocarbon sink. This significantly changes the behavior of the chemistry of the atmosphere leading to much higher levels of long-chain hydrocarbons. Detection of this change by the network analysis pipeline allows probing how network-based approaches can detect system-level changes in atmospheric chemistry. In these experiments, network metrics were able to distinguish between the perturbed and non-perturbed models for biotic atmospheres. This is likely due to the increased flux of $CH_3$ derived from the higher surface flux of $CH_4$, which is no longer being removed by conversion into $C_2H_2$ and then to haze. While the $CH_4$ concentrations were largely the same regardless of whether the model had been perturbed or not, the knock-on effects of increased $CH_3$ and loss of sinks led to considerable changes in the weighted topology of the network. This example clearly demonstrates how network topology can better reveal system-level differences. Additionally, the changes to the available reactions led to the presence of outliers not seen in the unperturbed distributions.

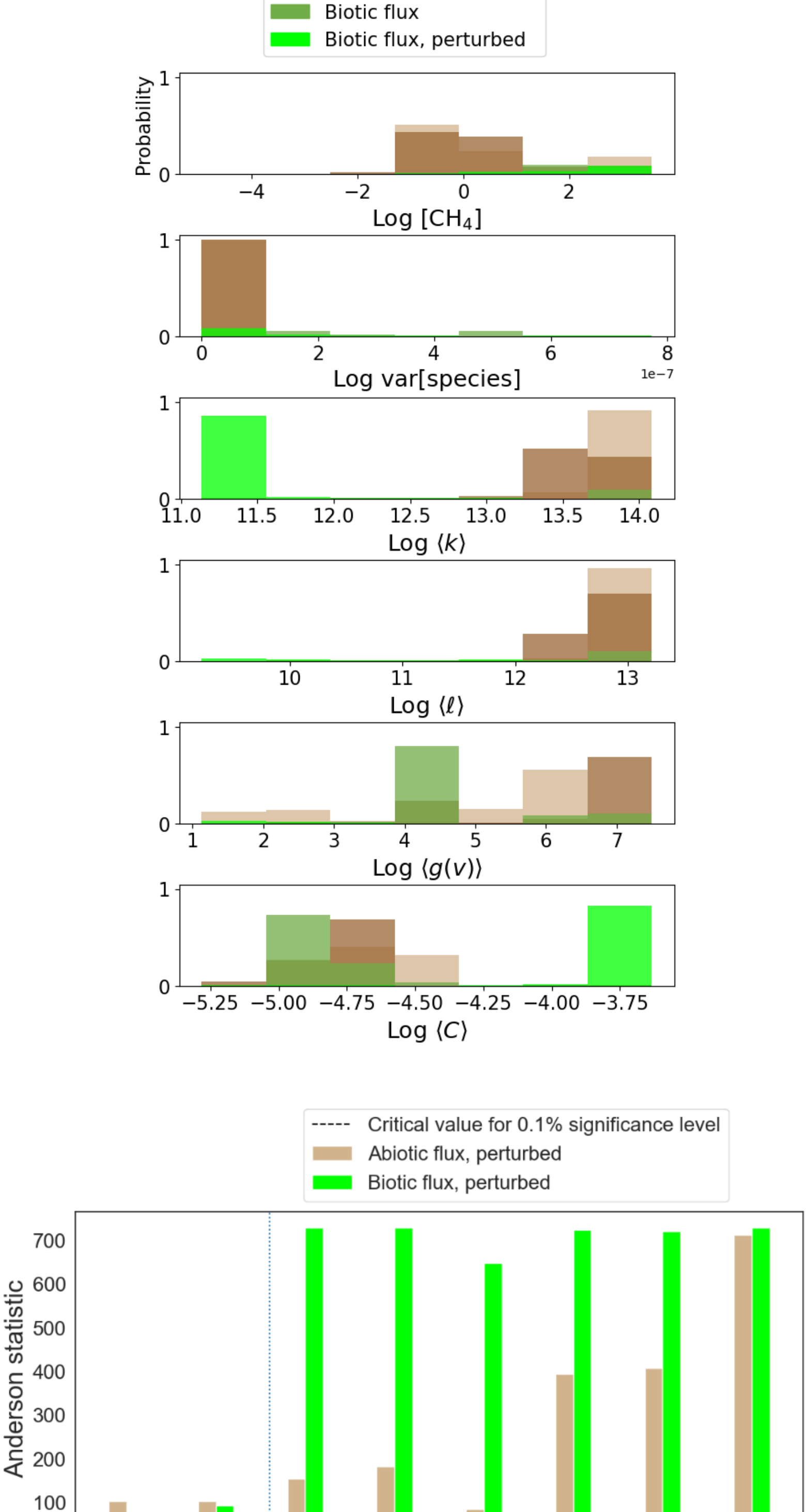

***Figure S5.*** *Top: Distributions of the probability (y-axis) for a simulated atmosphere to yield the given network metric value, global species variance or $CH_4$ abundance in log ppm (x-axis) over*

*each of the sampled Archean Earth datasets, probed at 0.2 mbar. Dark brown represents the abiotic flux model set; light brown represents the distributions after perturbing this model by removing* $C_2H_2$*. Dark green represents the biotic flux model set; light green represents the distributions after perturbing this biotic model. Bottom: Anderson scores of the biotic flux scenario as compared to the perturbed biotic model and the abiotic flux scenario compared to the perturbed abiotic model, demonstrating how the network metrics are much more sensitive to the perturbation than methane gas abundance or global species variance.*

## Bayesian Analysis

We integrated network metrics into a Bayesian framework to determine the posterior likelihood of life being the correct hypothesis given an observation, *O*:

$$P(life|O) = \frac{P(O|life) \times P(life)}{P(O)} \tag{12}$$

Probability distributions for our analyses were generated from our datasets using Gaussian kernel density estimation (KDE) – with the `gaussian_kde` package included in `SciPy` – to create an estimator for the kernel density from the distributions of each network metrics or $CH_4$ abundance (across 1000 points spanning the range of observed values). These were then used to create a normalized probability distribution for each network metric or gas abundance, where each atmospheric scenario was weighted equally and using a bandwidth parameter (between 0 and 1) scaled to each metric's standard deviation to smooth distributions accordingly. The normalized probability distributions were then implemented to calculate the conditional probabilities following Bayes' theorem.

*P(O|abiotic)* is the likelihood of a specific observation given *no life is present.* It was calculated by creating a Gaussian kernel density estimator of the value distribution of the given network metric from the abiotic models, using the `gaussian_kde` package included in `SciPy`. This estimator was then integrated over the value distribution, using the package's in-built `integrate_1D_box` function (to normalize to a probability between 0 and 1), with a bandwidth varying between 0.2 and 0.05 (depending on the metric being evaluated). This transforms the distribution of observed values into a probability distribution normalized between 0 and 1, *P(O|abiotic*), which can then be used as input for calculations using Bayes' theorem. The same process was repeated using the value distributions of the network metrics of the biotic models to calculate *P(O|life)*. These likelihood distributions were then used to calculate *P(O)* by summing the total probability for *any* observation by a known mechanism:

$$P(O) = P(O|abiotic) \times \left(1 - P(life)\right) + P(O|life) \times P(life) \quad (13)$$

*P(life|O)* was then evaluated for the network metrics, variance of species abundances and methane gas abundances, assuming values of *P(life)*=0.001, 0.01, 0.1, 0.5, and 0.9. The same process was repeated for both anomalous models where the distributions from the anomalous datasets were used in place of abiotic data.

## Posterior Distributions for Life

**Figures S6-S12** show probability distributions used in analysis of the posterior distribution of life for the different scenarios shown in **Figure 3** in the main text.

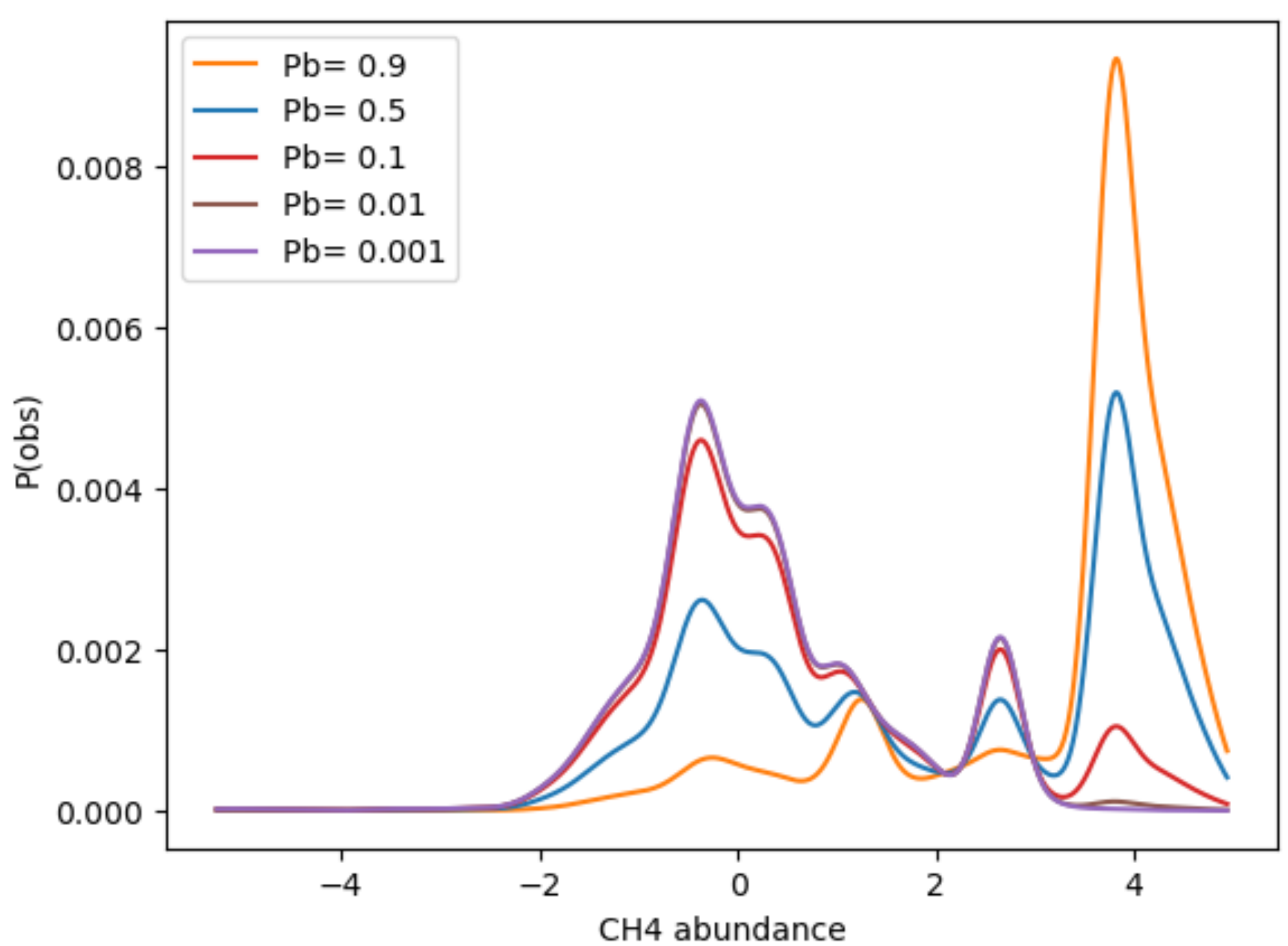

***Figure S6***. *P(observation) of given values of $CH_4$ abundance in log ppm.*

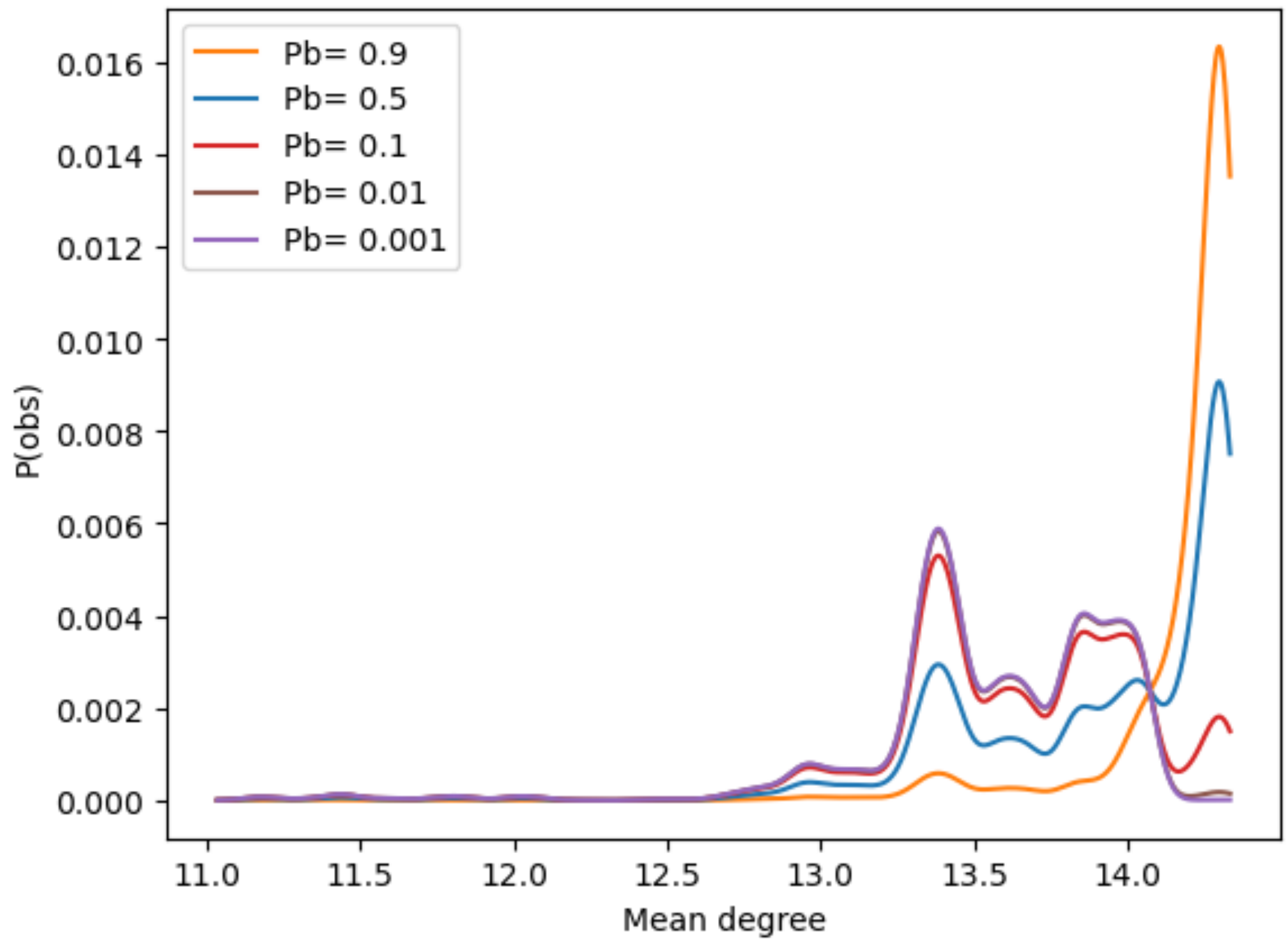

***Figure S7.*** *P(observation) of given values of log mean degree.*

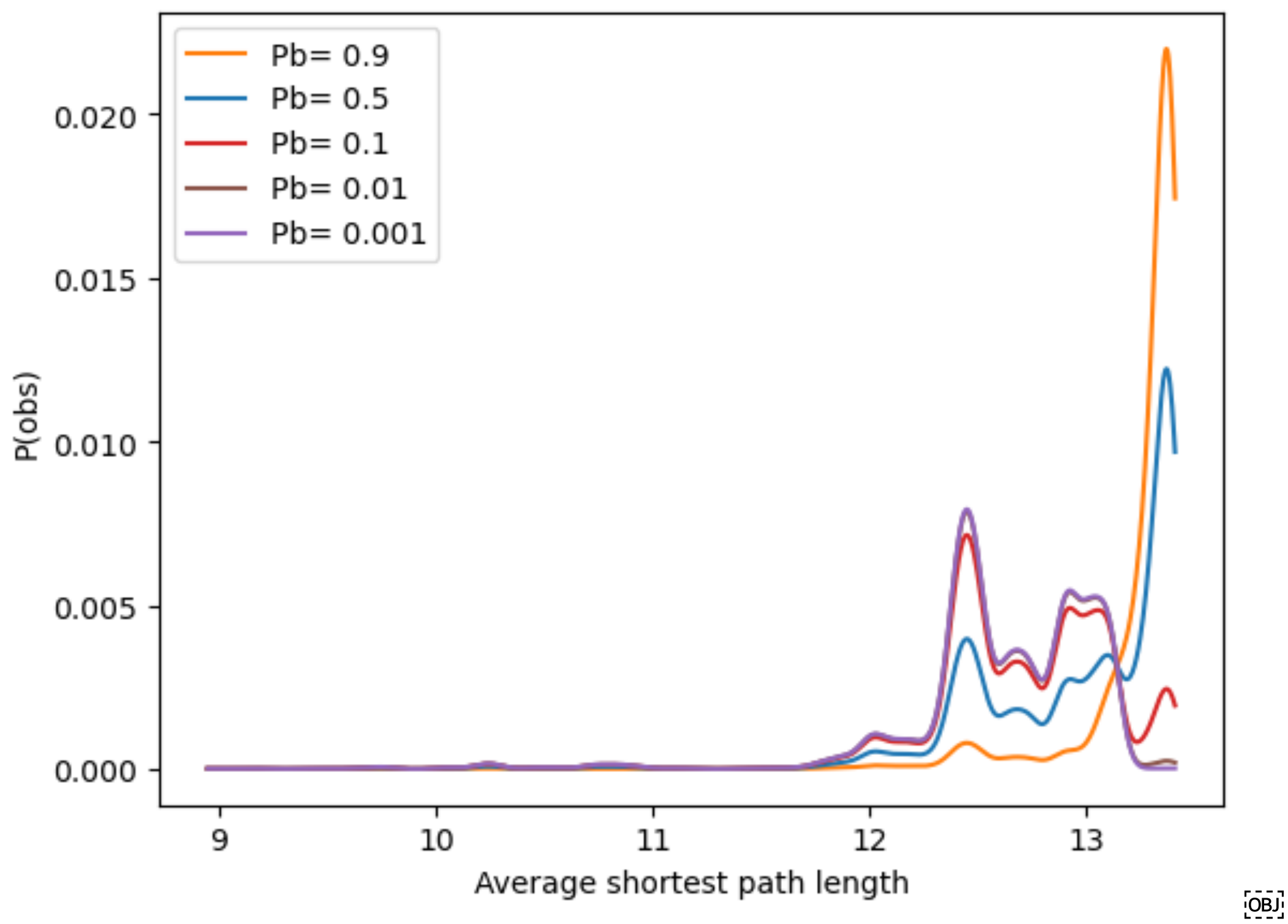

***Figure S8***. *P(observation) of given values of log average shortest path length.*

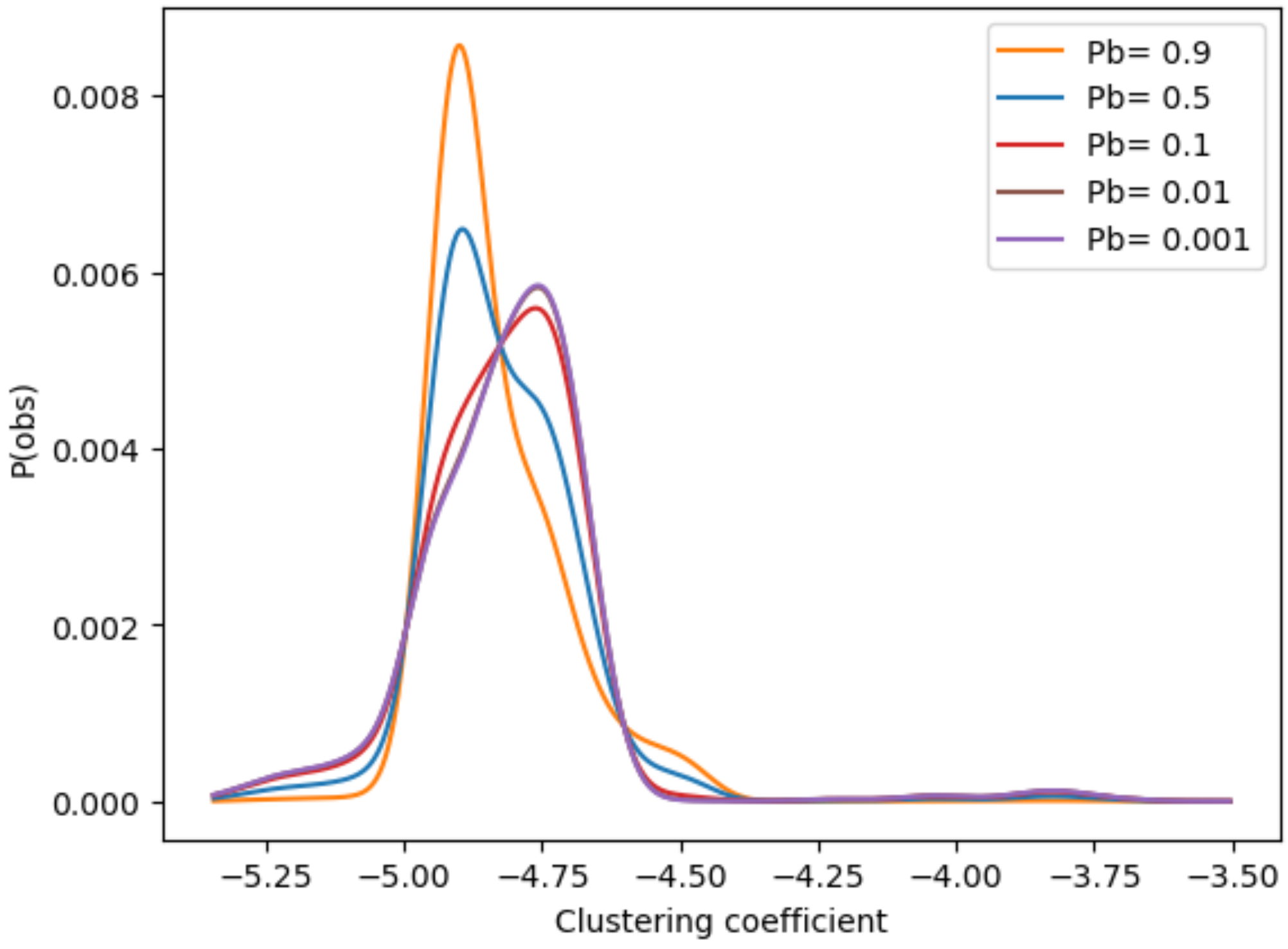

***Figure S9***. *P(observation) of given values of log average clustering coefficient.*

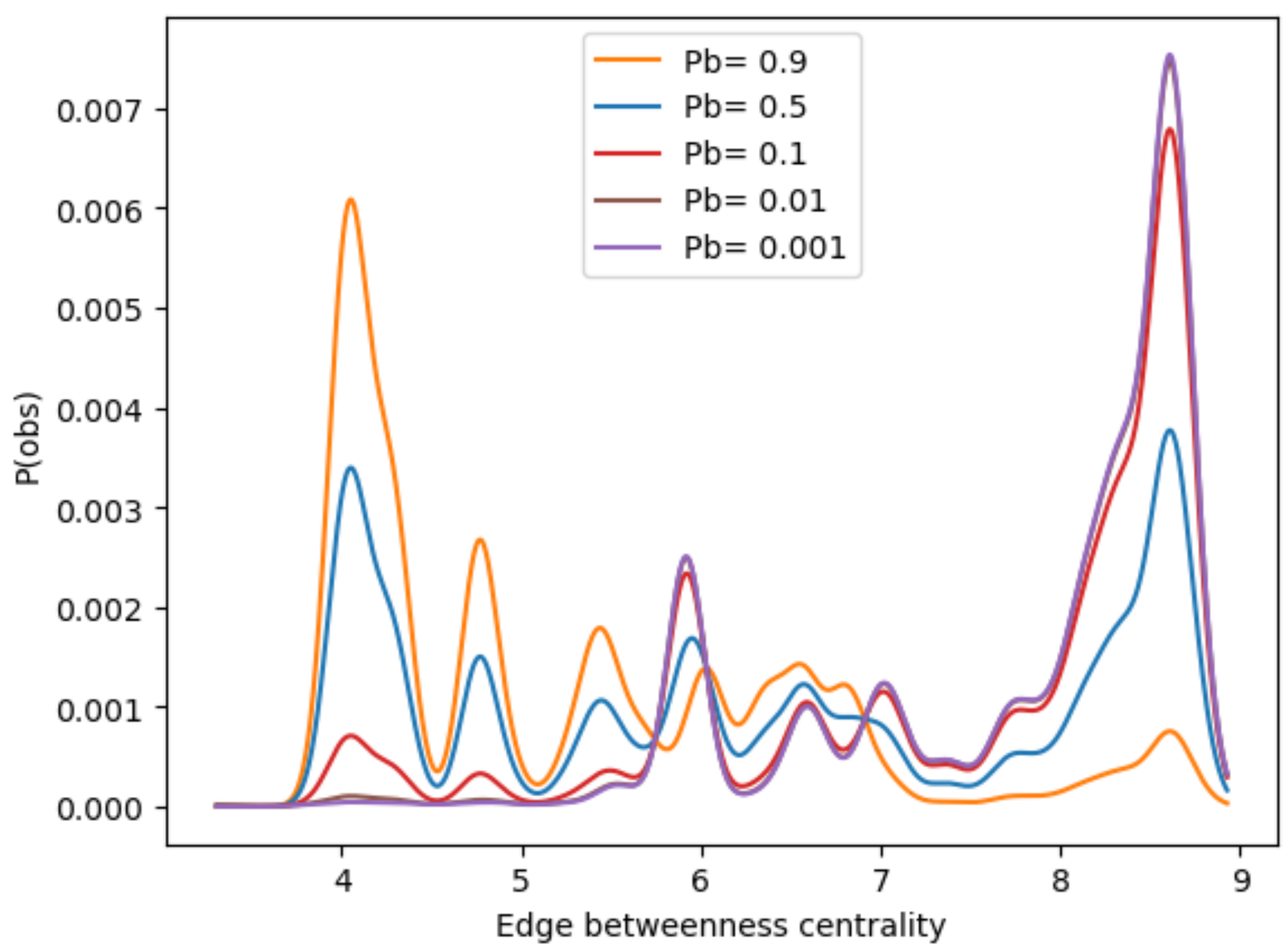

***Figure S10.*** *P(observation) of given values of log edge betweenness centrality.*

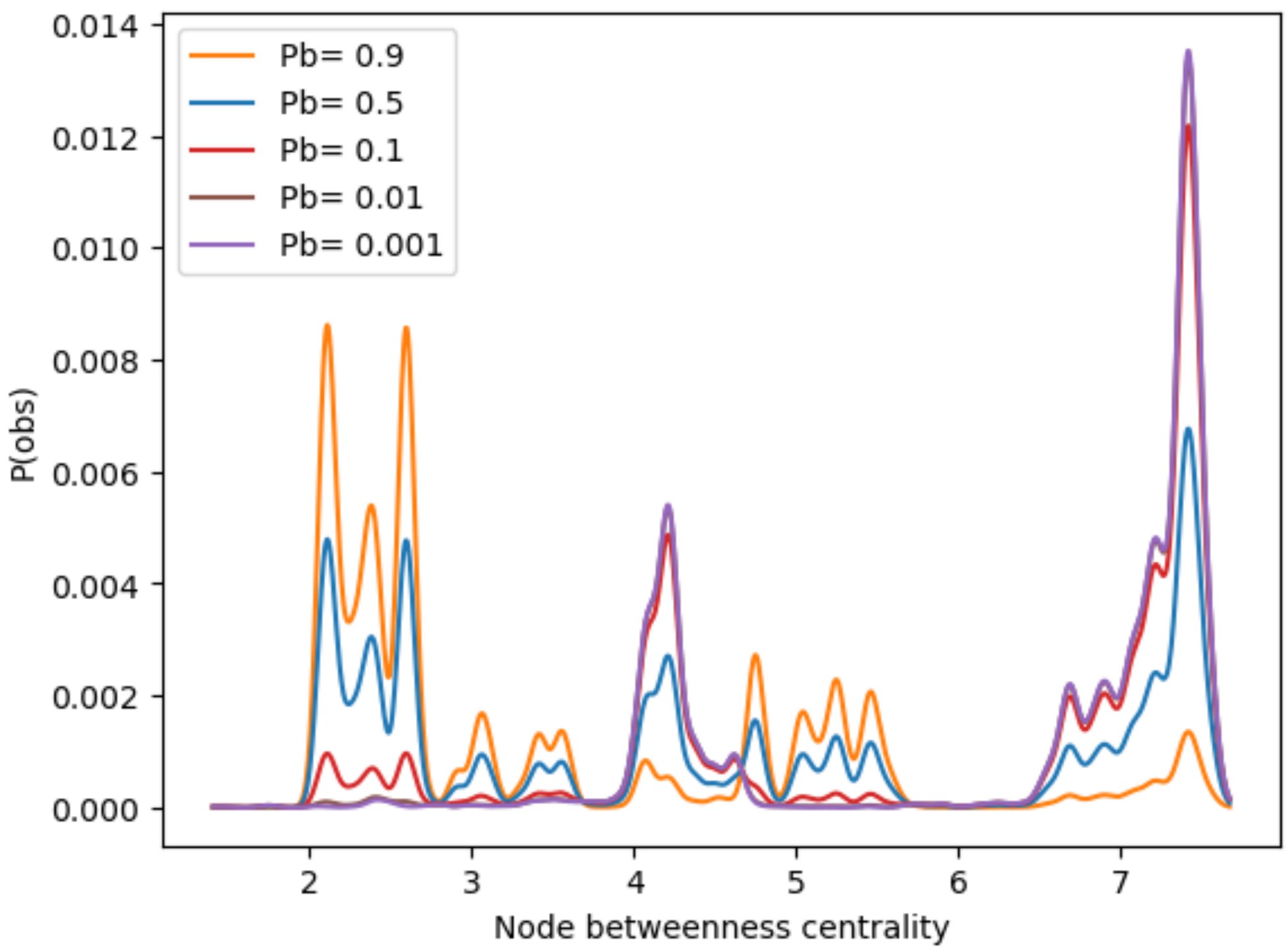

***Figure S11.*** *P(observation) of given values of log node betweenness centrality.*

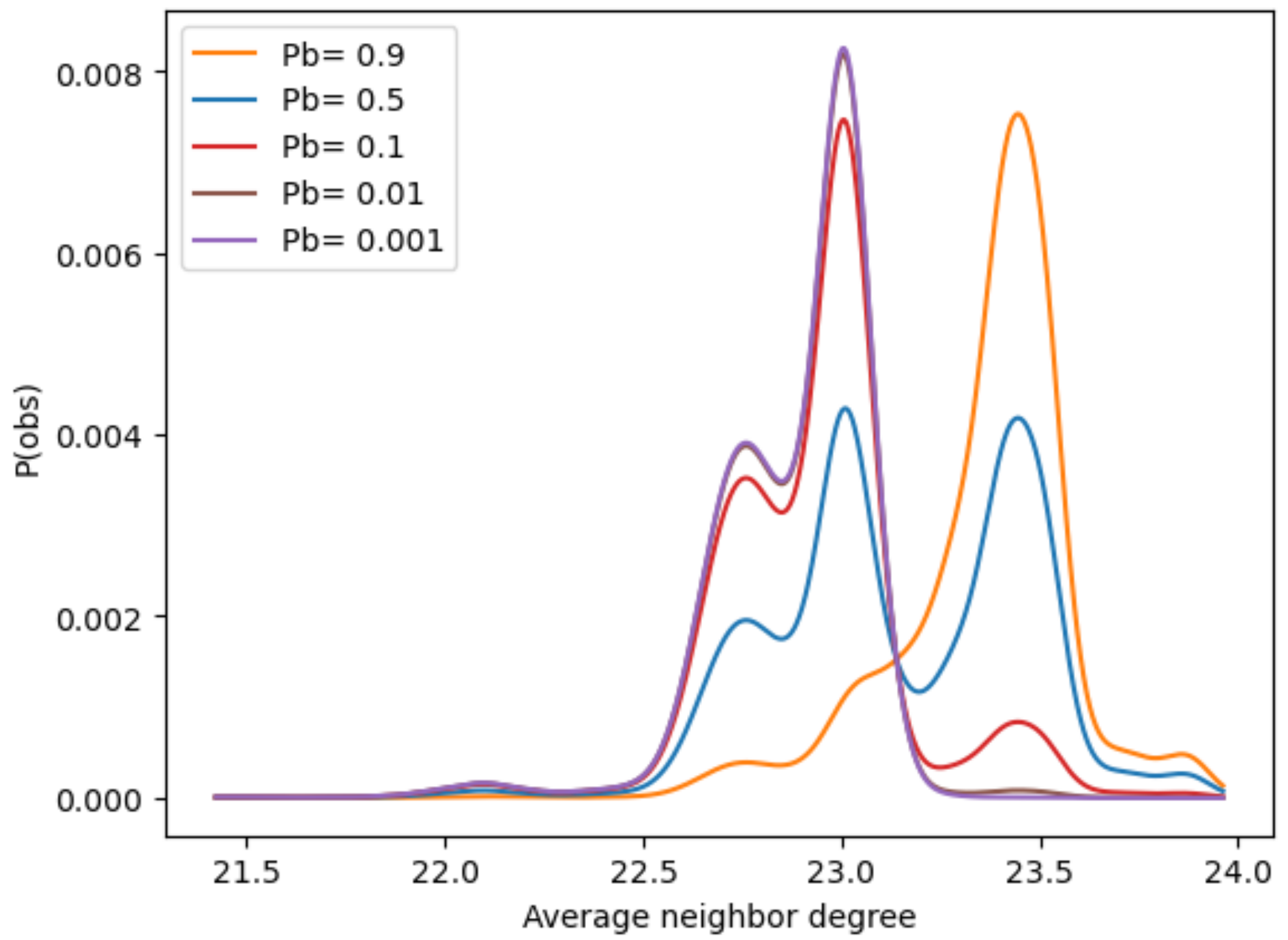

***Figure S12.*** *P(observation) of given values of log average neighbor degree.*

## Comparison between 0.2 mbar and 1 mbar Probed Altitudes

To test the impact of the chosen atmospheric altitude probed for abundances and CRN characteristics in our study (0.2 mbar or z = 64 km), we ran a comparative analysis where the relevant metrics are probed at an atmospheric pressure of 1 mbar (corresponding to z = 47 km). The results are presented in **Figures S13-S15** below, showing no significant difference between the two altitudes and speaking to the robustness of our approach, depending on what layer of the atmosphere is accessible to spectroscopic observations. The only noticeable difference concerns <*g*>, or average node betweenness centrality, which did not yield an interesting anti-biosignature window in our main analysis and does not in this case either.

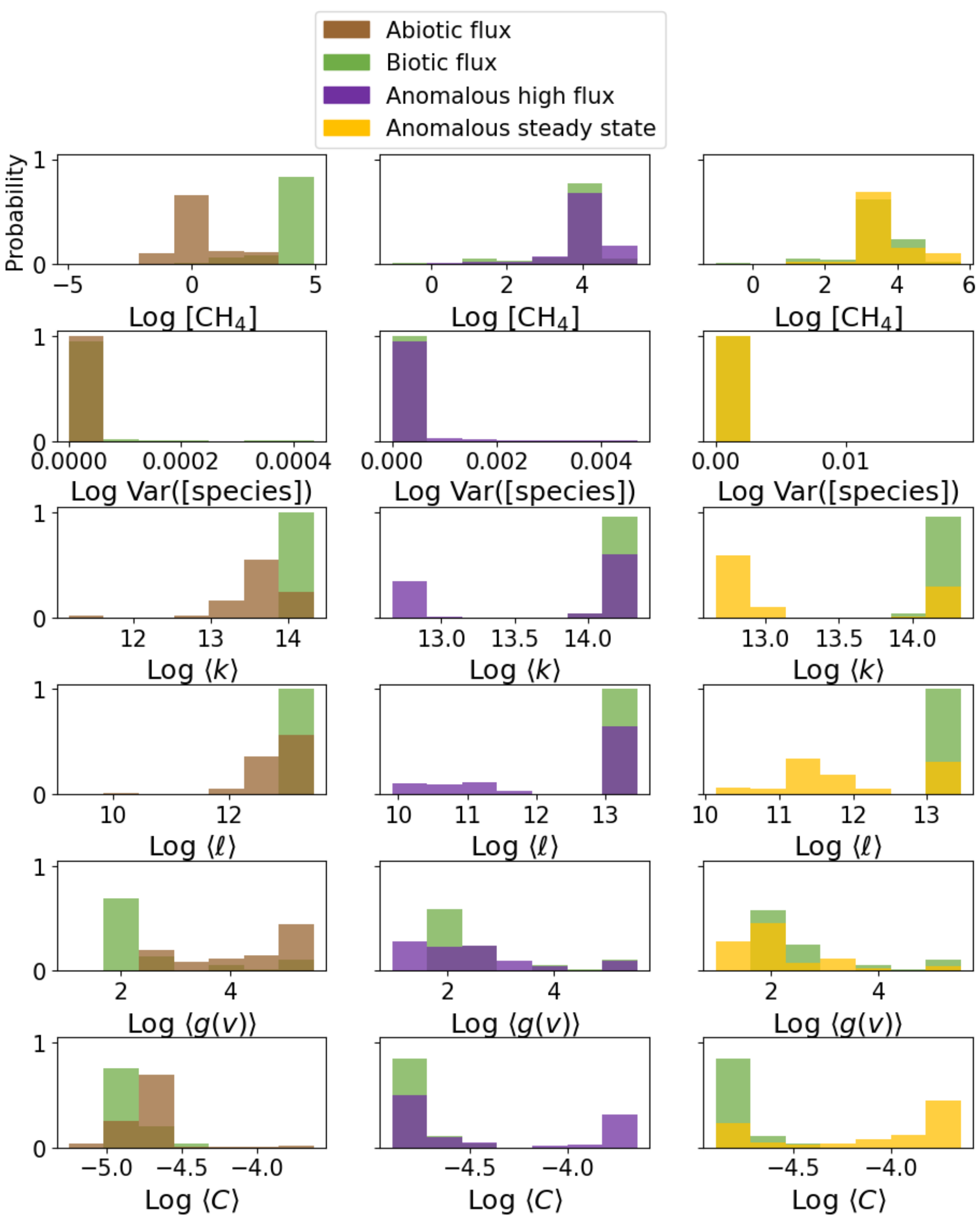

***Figure S13.*** *Distributions of the probability (y-axis) of a simulated atmosphere yielding the given network metric value, species abundance variance or* $CH_4$ *abundance in log ppm (x-axis) over each of the sampled Archean Earth datasets, probed at 1 mbar. Brown represents models where methane is modeled as an abiotic surface flux. Green represents models where methane is modeled as a biotic surface flux, with the methanogenesis pathway explicitly included in the chemical reaction network. Purple represents models where methane was modeled as originating from a surface flux of undetermined origin ("anomalous surface flux"). Yellow represents models where methane is modeled as an anomalous steady-state component of the planet's atmosphere with unknown source.*

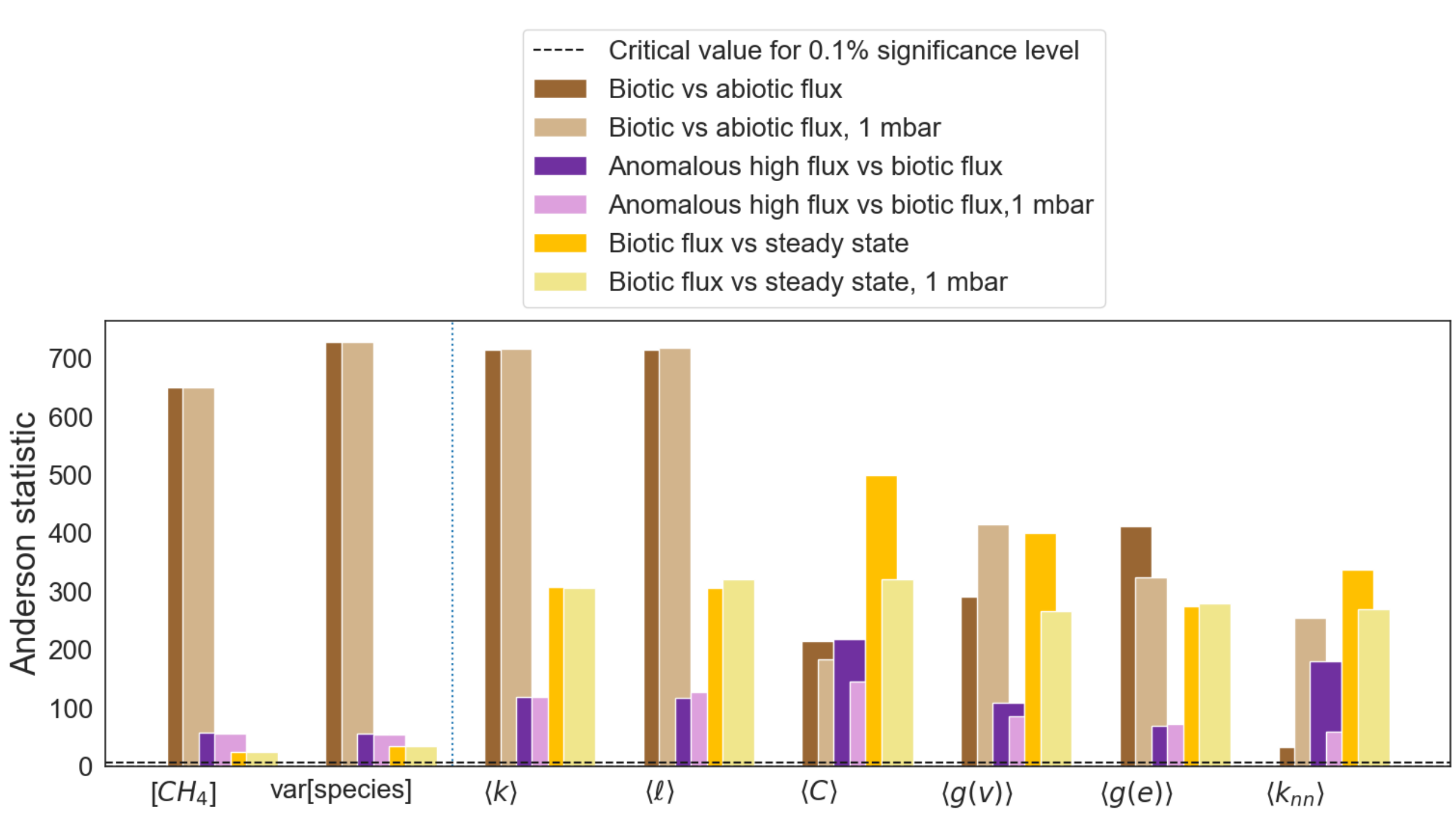

***Figure S14.*** *Anderson scores of the biotic flux scenario as compared to the three control models (abiotic flux, anomalous surface flux, and anomalous steady state), for CRN outputs probed at 0.2 mbar and 1 mbar (see Legend). Each bar indicates the performance of the corresponding metric in distinguishing between different modeled scenarios. The black horizontal dotted line indicates the critical value for distinguishability between model scenarios, at 0.1% significance level, and the blue vertical dotted line separates the networks metrics from* $CH_4$ *abundance and global species abundance variance.*

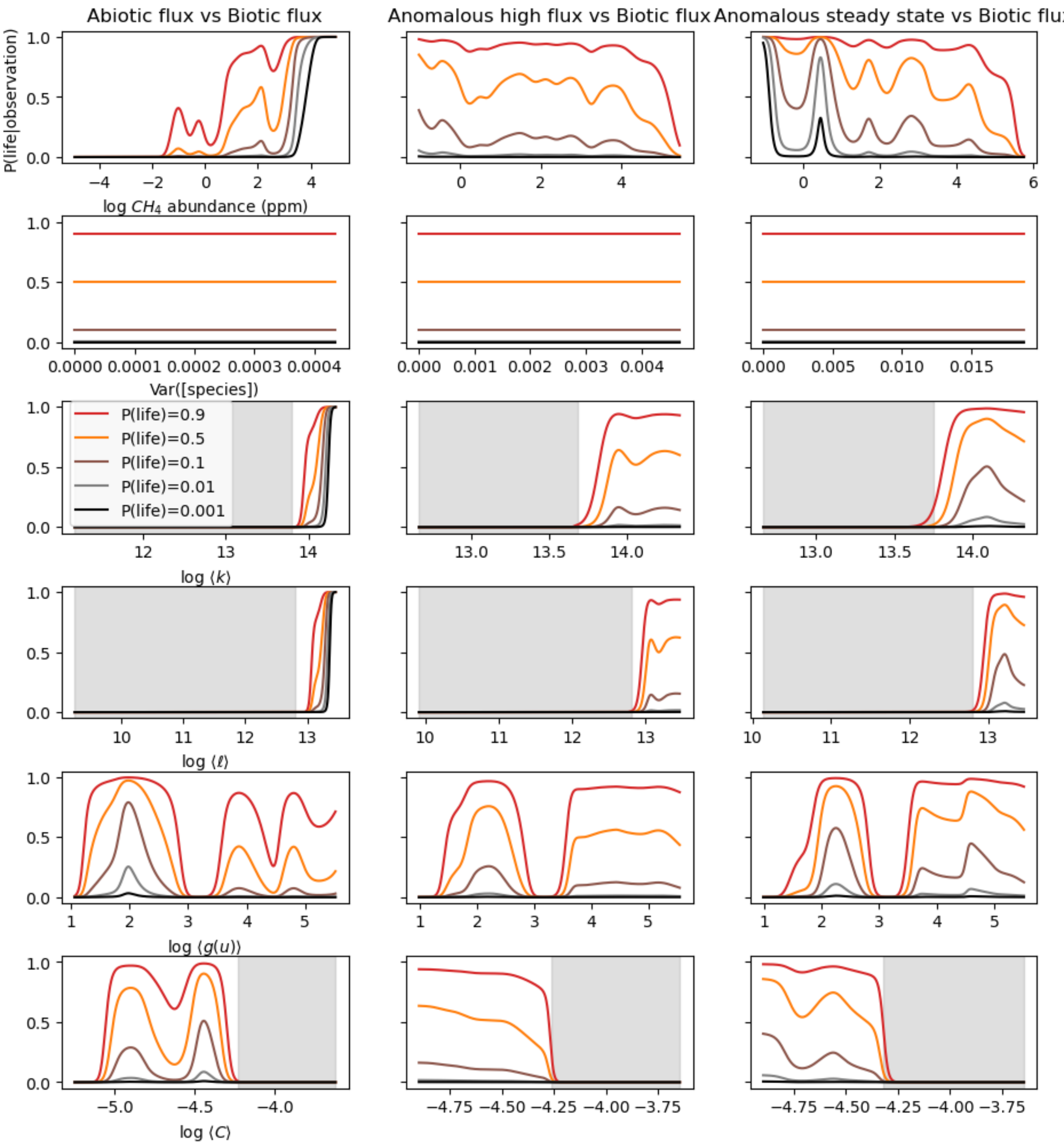

***Figure S15.*** *Bayesian analysis of the posterior likelihood of life, P(life|observation), assuming an observed value for network metrics and methane gas abundance, probed at 1 mbar. Given that the probability for life to exist on an exoplanet is unknown, we assess the posterior likelihood over a range of prior probabilities for life, P(life). Grey regions highlight metric values that are associated with a zero probability of life and are therefore anti-biosignatures for the sets of models compared herein. In general, the subset of network metrics studied here can provide greater constraints on the probability of life given a set of observations than $CH_4$ abundance alone, allowing distinguishing cases where Earth-like methanogenesis can be ruled out even when $CH_4$ abundance alone is not conclusive.*

## Modern Earth-like Atmospheres with and without Technologically Produced CFCs

To test the impact of CFC emissions on atmospheric CRN topology, we used the existing dataset generated by the NASA Frontiers Development Lab using `PyAtmos`. As a test of the capabilities of `PyAtmos`, FDL modeled a large set of conditions with a resulting dataset composed of over 120,000 modeled Modern-Earth analogue atmospheres with varying amounts of $CH_4$, $CO_2$, $H_2$, and $O_2$ (see **Table S1**). All cases are characterized by the same temperature-pressure profile, shown in **Figure S16**. From this dataset, we randomly selected 5,000 models and re-ran them after incorporating a surface flux of $CCl_2F_2$, better known as CFC-12 or Freon™; this gave us a total of 10,000 models of modern Earth-like atmospheres, half with CFC-12, half without. CFC-12 is notorious for its depleting effect on atmospheric ozone; when exposed to light, it gives rise to a Cl radical:

$CCl_2F_2 + hv \rightarrow CClF_2 + Cl$

This radical can in turn catalyze the conversion of $O_3$ into other oxygen species, via

$Cl + O_3 \rightarrow ClO + O_2$

or

$ClO + O_3 \rightarrow CLOO + O_2$

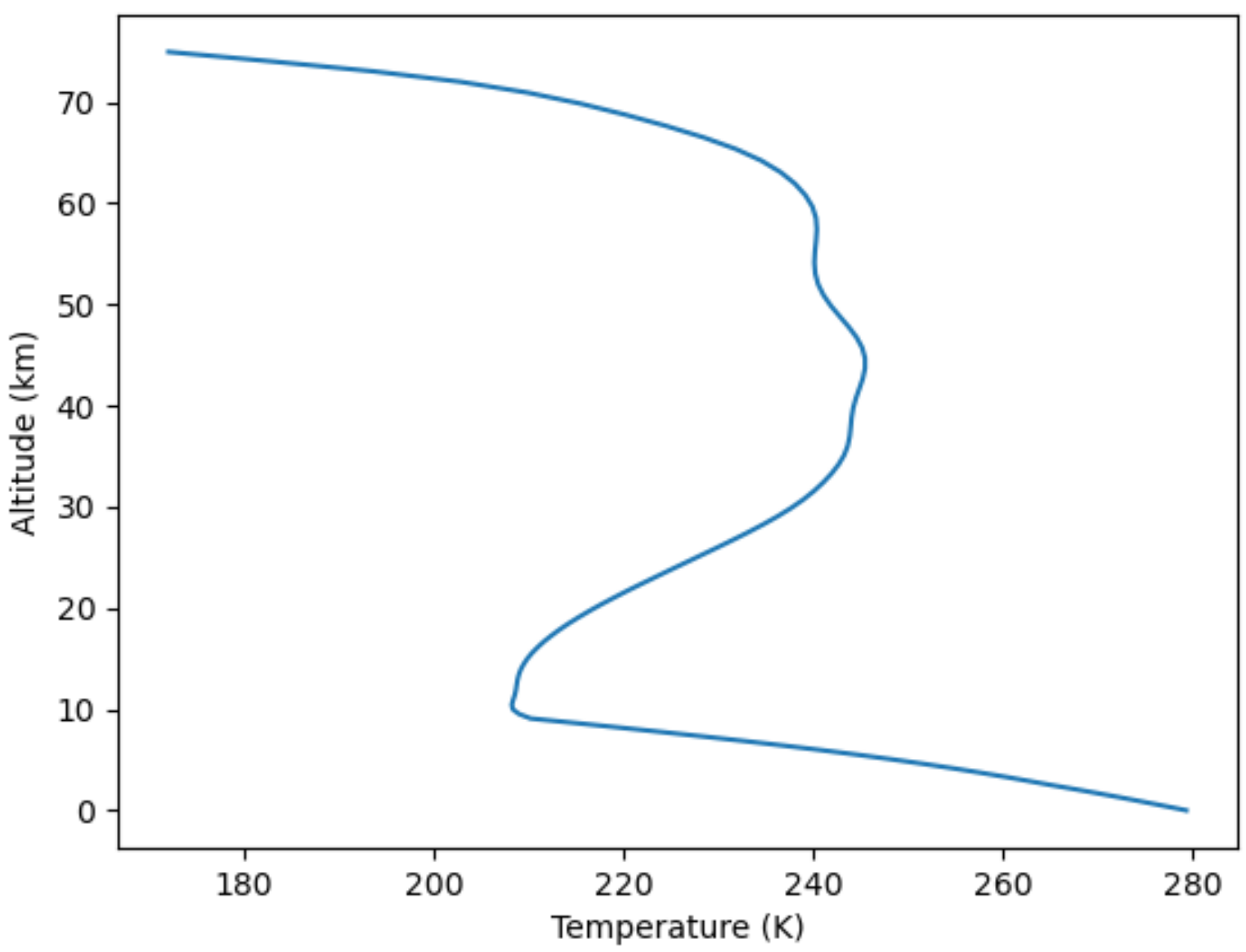

***Figure S16:*** *The temperature profile as a function of scale height, input for the Modern-Earth simulations, adopted from the NASA Frontiers Development Lab. All Modern Earth model scenarios use the same temperature-pressure profile in this study.*

To simplify atmospheric chemistry to constrain detectability, CFC-12 abundances were modeled as a steady state, with values ranging from 0.05 to 5 ppb, based on the trajectory of the abundance of industrial CFC-12 in the Earth's atmosphere as modeled by Haqq-Misra et al (2022).

| Species | Concentration ranges (fractional) |
|---|---|
| $CH_4$ | $1.63\times10^{-6}$-0.13 |
| $CO_2$ | $8\times10^{-8}$-0.4 |
| $H_2$ | $6\times10^{-8}$-0.19 |
| $O_2$ | 0.02-0.5 |
| CFC-12 | $4.2\times10^{-11}$-$2.3\times10^{-6}$ |

***Table S2.*** *Species concentrations that are free parameters in the modern Earth analogue population of models*

In the resulting networks, we observe that the network with CFC-12 emissions indeed gains 5 nodes corresponding to CCl2F2, CCIF2, Cl, ClO and CLOO, and 14 edges – 2 connected to CCl2F2 and 12 to the Cl radical.